\documentclass[5p, preprint, 
]{elsarticle}

\usepackage[dvipdfmx]{hyperref}
\usepackage{amsmath,amssymb,amscd,amsbsy,amsgen,amsopn,amstext,amsxtra,bm}
\usepackage{dblfloatfix} %

\usepackage{caption}
\usepackage[subrefformat=parens]{subcaption}

\usepackage{siunitx}

\usepackage{here}

\journal{Journal of \LaTeX\ Templates}

\begin{document}

\begin{frontmatter}

\title{The 3-dimensional Fermi liquid description for the normal state of cuprate superconductors}


\author[mymainaddress]{Setsuo Misawa\corref{mycorrespondingauthor}}
\cortext[mycorrespondingauthor]{Corresponding author}
\ead{fermild@gmail.com}


\address[mymainaddress]{Professor emeritus, Department of Physics, College of Science and Technology, \\
Nihon University, Kanda-Surugadai, Tokyo 101-8308, Japan}

\begin{abstract}
The quasiparticles in the normal state of cuprate superconductors have been shown to behave universally as a 3-dimensional Fermi liquid.
Because of interactions and the presence of the Fermi surfaces (or Fermi energies), the quasiparticle energy contains, as a function of the momentum $\bm{p}$, a term of the form $(p-p_0)^3  \ln ( | p-p_0 | / p_0 )$, where $ p  = | \bm{p} |$ and $p_0$ is the Fermi momentum.
The electronic specific heat coefficient $\gamma(T)$, electrical resistivity, Hall coefficient and thermoelectric power divided by temperature $T$, follow the logarithmic formula $ a - b T^2 \ln ( T/T^*) $, $a$, $b$, and $T^*$ being constant.
Singularities in the Landau $f$-function produce the $T^2 \ln T$ dependence of the magnetic susceptibility $\chi (T)$, and Knight shift, which gives rise to the phenomenon of the susceptibility maximum.
The logarithmic $T$-dependence of the transport properties arises exclusively from the impurity scattering in 3-dimensional (3D) systems, but does not from the electron-electron scattering in 2D systems.
The above logarithmic formula has been shown to explain universally the experimental data for the normal state of all cuprate superconductors.
The decrease of $\gamma(T)$ or $\chi(T)$ with decreasing $T$ is not due to the appearance of pseudogap or spin gap but due to its $T^2 \ln T$ variation.
\end{abstract}

\begin{keyword}
\texttt{
transport properties,
logarithmic temperature dependence,
impurity scattering,
susceptibility maximum,
kink phenomenon
}
\end{keyword}

\end{frontmatter}

\section{Introduction}

In the study of high $T_{\mathrm{C}}$-cuprates, the most serious problem is to clarify the anomalous properties of the normal state.
To explain the anomalous temperature dependence of transport properties and to seek for the origins of the pseudogap, spin gap, kink phenomena and so on are the most important concerns of this system.
In the literature, these anomalies are usually attributed to the effect of the non-Fermi liquid or anomalous metals.

Anomalous properties similar to the above also appear in the normal state of iron-based superconductors;
these anomalies have been clarified by the present author or the basis of 3D normal Fermi liquid \cite{1,2}.
Concerning cuprate superconductors, in 2003, we have already published a preliminary report entitled ``strong evidence for the 3-dimensional Fermi liquid behaviour of quasiparticles in high-$T_{\mathrm{C}}$ cuprates'' \cite{3}.

Here, in this paper, we shall present a unified description in which all anomalies in the normal state of high-$T_{\mathrm{C}}$ cuprates can be systematically explained in terms of a normal 3D Fermi liquid.

The basic elements of high-$T_{\mathrm{C}}$ cuprates, e.g. electrons and holes in $\mathrm{La}_{2-x} \mathrm{Sr}_{x} \mathrm{Cu} \mathrm{O}_4 $(LSCO), are placed on the 2-dimensional $\mathrm{CuO_2}$ plane.
It does not mean, however, that the state of these particles is 2-dimensional.
The band-structure theory predicts on anisotropic 3D metals for any $x$ \cite{4}. 
Concerning this dimensionality, the results of electrical resistivity, $\rho_{ab}$ and $\rho_c$, measurements for LSCO single crystals are instructive \cite{5}.
For $x=0.10$ sample, $\rho_c$ increases almost exponentially, as for insulators, with decreasing temperatures for $T \lesssim 200 \; \mathrm{K} $.
For $4$ samples, $0.12 \leq x \leq 0.30$, however, the electric conduction becomes metallic and the resistivity follows the 3D formulae.
The dimensionality of the system can also be judged by the specific heat coefficient. 
As shown in Sec.~2, it can be definitely concluded that quasiparticles in LSCO can never be 2-dimensional, but be 3-dimensional.

We here stand on the basic principle that the quasiparticles in the normal state of high $T_{\mathrm{C}}$-cuprates constitute a 3-dimensional Fermi liquid of Landau's spirit.
The Fermi liquid is characterized by both the interaction between particles and the presence of Fermi surfaces; mathematically, the discontinuity of the Fermi distribution function in 3D naturally produces logarithmic functions in physical quantities.

In underdoped cuprates, however, Fermi spheres are usually incomplete, but compose Fermi arcs.
Concerning this, Yoshida et al. \cite{6'} have studied the spectra of angle-resolved photoemission experiments for highly doped LSCO samples to find that sharp quasiparticle peaks crossing the Fermi energy do exist; the discontinuity around the Fermi energy really survives.
Therefore, our starting point is that, at absolute zero, the Fermi distribution function is $1$ for quasiparticles below the surface, and is zero for those beyond that; non-analytic, i.e. logarithmic, functions arise exclusively from this discontinuity.
Nevertheless, we assume that, at absolute zero, the Fermi distribution function is $1$ for quasiparticles below the surface, and is $0$ for those beyond that;
all results which are derived on this assumption concerning the temperature dependence of thermodynamic and transport properties are fully consistent with the experimental data.

In 1958, Galitsky \cite{6} has shown that the energy of the quasiparticle with the momentum $p$ contains a term $( p - p_0 )^3 \ln | p-p_0 |$, $p_0$ being the Fermi momentum.
In accordance with this, Migdal \cite{7} has shown, on the basis of the Green function theory, that the real part of the self-energy part,  $\Sigma(\varepsilon)$, contains the $\varepsilon^3 \ln | \varepsilon |$ term, where $\varepsilon$ is the energy of the quasiparticle measured from the Fermi energy $\mu$.
We shall show that this $\varepsilon^3 \ln |\varepsilon|$ dependence has been experimentally observed in the kink phenomena.

The above $\varepsilon^3 \ln |\varepsilon|$ term produces the $\varepsilon^2 \ln |\varepsilon|$ term in the density-of-states function $\nu(\varepsilon)$, and all physical quantities relating with $\nu (\varepsilon)$ show the logarithmic temperature dependence.
The electronic specific heat coefficient $\gamma (T)$ is given by $\gamma_0 + \gamma_1 T^2 \ln ( T/ T^{*}_{\gamma})$, where $\gamma_0$, $\gamma_1$, and $T^{*}_{\gamma}$ are constant.

Since the electrical resistivity due to the impurity scattering depends on $\nu (\varepsilon)$, the $T$-dependence of the resistivity is given by $\rho (T) = \rho_0 + \rho_1 T^2 \ln ( T/ T^*_{\rho} )$, where $\rho_0$, $\rho_1$ and $T^*_{\rho}$ are constant;  even if the Umklapp Coulomb scattering is included, this form of $\rho(T)$ is not altered.
The Hall coefficient $R_{\mathrm{H}} (T)$ being given by a functional of resistivity functions arising from multiple Fermi surfaces, of electrons and/or holes, it varies with $T^2 \ln T$; for higher temperatures, $T^4 ( \ln T )^2$ and $ T^4 \ln T $ terms become effective for $R_{\mathrm{H}} (T)$.
Since the thermoelectric power $Q(T)$ is proportional to the average of $\varepsilon$, it is given by the form $ q_0 T  + q_1 T^3 \ln ( T/ T^*_Q ) $, $q_0$, $q_1$ and $T^*_Q$ being constant.

The temperature dependence of the magnetic susceptibility (or Knight Shift) arises from spin-antisymmetric part of the Landau $f$-function; $g(p,p')$.
Near the Fermi surface, $g(p,p')$ contains terms of the form $(p-p_0)^2 \ln |p-p_0| $; at finite temperature this function produces the $T^2 \ln T$ dependence of $\chi (T)$; $\chi(T) = \chi_0 - b T^2 \ln ( T/ T^*_{\chi} )$, where $\chi_0$, $b$ and $T^*_{\chi}$ are constant.
From this we predict the universal law of the susceptibility maximum.
Since $b>0$, all paramagnetic electron system should exhibit the maximum at $T_{\mathrm{max}} = T^{*}_{\chi} / \sqrt{ e } \fallingdotseq 0.607 \, T^{*}_{\chi} $.

Here it is important to note that the so-called spin fluctuation theory (SFT) \cite{9} or its related theories cannot predict the $T^2 \ln T$ dependence of $\chi (T)$ and the susceptibility maximum.
In SFT, the free energy of the system is expressed in terms of frequency $(\omega)$ and wave-number $(k)$ dependent dynamical susceptibility $\chi(k,\omega)$.
$\chi(k,\omega)$ contains logarithmic functions such as $\ln \{ 1- s \pm ( \tilde{k} / 2 ) \}$, where $ \tilde{k} = k/ k_0$, $k_0 = p_0 / \hbar, s= \omega/ k v_0$.
Here, most authors expand this in powers of $s$ and $\tilde{k}$ to yield
\[
\chi (k,\omega) \propto 1- A \tilde{k}^2 + i B s \: \: ,
\]
$ A $ and $B$ being constant.
This process inevitably eliminates the effect of the discontinuity in the Fermi distribution function, and no $T^2 \ln T$ term appears.

The pseudogap problem raises, for this origin, various assertions among various authors.
It should be remarked that Deisz et al \cite{10} have evaluated quasiparticle properties in the fluctuation exchange approximation on the half-filled Hubbard model.
By calculating the self energy on the basis of a single boson (spin or charge fluctuation) exchange, they have found a clear dip in the density-of-states curve $\nu(\varepsilon)$ around the Fermi energy.

Loram et al \cite{11} have attributed the origin of the pseudogap to the $| \varepsilon |$ dependence of $\nu(\varepsilon)$ which should give rise to the $T$-linear term in $\gamma(T)$ and $\chi(T)$.
Their experimental data for $\gamma(T)$, however, do not show such dependence; for the $x = 0.08$ sample of $\mathrm{ La_{2-{\it x}} Sr_{\it x} Cu O_{4} }$, $50 \mathrm{K} \lesssim T \lesssim 125\, \mathrm{K}$,
for example, the specific heat is shown to obey $\varDelta \gamma = \gamma(T) - \gamma(0) \propto T^{0.64} $.
Later, we shall show that their data follow more precisely the $T^2 \ln T$ law for wider temperature range.

The spin gap problem has been raised since the temperature dependence of the inverse relaxation time $1/ T_1 T$ shows a peak structure in $\mathrm{YBa_2 Cu_4 O_8}$ \cite{12'}; the origin of the maximum has not been clarified.
Concerning the $T$-dependence, both $\chi (T)$ and $ 1/ T_1 T $ are correlated with the logarithmic singularities appearing in the dynamical susceptibility $ \chi ( k, \omega ) $; analytically rigorous evaluation of $1 / T_1 T$ has not been performed yet. 
Experimentally, however, Alloul et al. \cite{13'''} have found that, in $ \mathrm{YBa}_2 \mathrm{Cu}_3 \mathrm{O}_{6+x} $ samples, $(T_1 T)^{-1/2}$ and $ K(T) $ satisfy the Korringa relation, and Horovatic et al. \cite{14'''} have observed that, in $ \mathrm{YBa}_{2} \mathrm{Cu}_{3} \mathrm{O}_{6.52} $ single crystal, $1/T_1 T $ scales linearly with $K(T)$; 
these proves that, at low temperatures, $1/T_1 T$ varies as $T^2 \ln T$ in accordance with $K(T)$ or $ \chi (T)$, and that the  $ 1/ T_1 T $ maximum is  universally observed. 
When the maximum temperature becomes lower than the superconductivity transition temperature $T_{\mathrm{C}} $, the maximum disappears.

Here we should mention the theoretical treatment by Kontani et al.~\cite{14'} on the basis of nearly antiferromagnetic, 2-dimensional (2D) Fermi liquids.
In order to explain ``so-called'' non-Fermi liquid behaviours, they have seriously treated the vertex corrections to satisfy the conservation laws and found that the Hall coefficient $R_{\mathrm{H}}$ is positive for hole-doped compounds and negative for electron-doped compounds.
Their results, however, do not reproduce the experimental data for temperatures below $500 \; \mathrm{K}$ \cite{12}.
As will be seen in Sec.~7, for explaining the whole range of $T$, it is of critical importance to consider, instead of the electron-electron scattering in 2D systems, the impurity scattering in 3D systems. 

This paper is organized as follows:
As representatives of high-$T_{\mathrm{C}}$ cuprates, we treat the normal state of $\mathrm{ La_{2-{\it x}} Sr_{{\it x}} }$ $\mathrm{Cu O_4} \mathrm{ ( LSCO) }$, $\mathrm{ YBa_{2} Cu_{2} O_{7-{\it x}}}$ $\mathrm{ (YBCO) }$ and $\mathrm{Nd_{2- {\it x}} Ce_{\it x} Cu } $ $ \mathrm{O_4 }$.
In Sec.~2, the electronic specific heat coefficient is derived on the basis the quasiparticle energy, and the proof for the system not being 2-dimensional is presented.
In Sec.~3, the relation between the kink phenomenon and the $\varepsilon^3 \ln \varepsilon $ dependence of the self energy is discussed.
In Sec.~4, the temperature dependence of the magnetic susceptibility and the phenomenon of the susceptibility maximum are presented;
the temperature dependence of the nuclear spin-lattice relaxation time is discussed in connection with its maximum.
In Sec.~5, we deal with the unified explanation for the temperature dependence of various transport properties.
In Sec.~6,~7, and~8, we analyse existing experimental data of the electrical resistivity, Hall coefficient and thermoelectric power to show that all data can be explained on the basis of 3D Fermi liquid model.
Sec. 9 will be devoted to conclusions.

\section{Eletronic Specific Heat Coefficient of LSCO; the Proof for the System not being 2-dimensional}

The entropy $S$ of an interacting fermion system is given, according to Landau, as a functional of the quasiparticle distribution function $n$, by
\begin{align} \label{1.1}
S = - k_{\mathrm{B}} \int d \varepsilon \; \nu (\varepsilon)  \, \Big\{ \, n \ln n + (1-n) \ln (1-n) \Big\} \: .
\end{align}
At equilibrium, $n= ( e^{\beta \varepsilon} + 1 )^{-1} $, $\beta = 1/k_{\mathrm{B}} T $, $k_{\mathrm{B}}$ and $\nu (\varepsilon )$ being the Boltzmann constant and the density-of-states function.

The quasiparticle energy for an interacting fermion system was first obtained by Galitsky \cite{6} in terms of the scattering length for two quasiparticles, $a$;
\begin{align} \label{1.2}
 \varepsilon ( p )  = \varepsilon ( p_0 ) + \mathrm{O} ( \tilde{p} , \tilde{p}^2 , \tilde{p}^3 ) - \lambda \tilde{p}^3 \ln | \tilde{p} | \: ,
\end{align}
where $ \tilde{p} = ( p - p_0 ) / p_0 $, and $\mathrm{O} ( \tilde{p}, \tilde{p}^2 , \tilde{p}^3 )$ denotes a sum of terms of the orders $\tilde{p}$, $\tilde{p}^2$, and $\tilde{p}^3$; $\lambda = ( 4/3 \pi^2 )(p_0^2 / m ) ( k_{0} a )^2$ represents the interaction strength of the system, where $m$ is the mass of a particle, and $k_{0} = p_0 / \hbar $ is the Fermi wave number.
From this, the density-of-states function for both spins is derived as
\begin{align} \label{1.3}
\nu (\varepsilon)
	&= \frac{  8  \pi  p^2  }{  (  2 \pi  \hbar  )^3  }
		\left( \frac{ \partial \varepsilon}{ \partial p } \right)^{-1}
\notag
\\
	&= \frac{ m  p_0 }{  \pi^2  \hbar ^3 }
		\left\{  1  +  \mathrm{O}( \varepsilon, \varepsilon^2)
			+ \left( \frac{ k_0 a }{  \pi  }  \right)^2
				\left( \frac{  \varepsilon  }{  \varepsilon^*_0  }   \right)^2
				\ln \frac{  | \varepsilon |  }{  \varepsilon^{*}_{0}  }
				\right\} \: ,
\end{align}
where $\varepsilon^{*}_{0} = p_0^2 / 2m^{*}$, $m^{*} = p_0 / ( \partial \varepsilon / \partial p )_{p_0}$ is the effective mass of a quasiparticle.
$S$ can be calculated by Eqs. \eqref{1.1} and \eqref{1.3}, where the logarithmic divergence arising in the asymptotic expansion can be survived by considering the effect of finite temperatures.
Thus the specific heat coefficient $\gamma (T) = \partial S / \partial T $ is obtained, up to order $T^2$, as
\begin{align} \label{1.4}
\gamma(T)
	& = \gamma_0  + \gamma_1  T^2 \ln T + \gamma_2 T^2
\notag \\
	&= \gamma_0
		+ \gamma_1 T^2 \ln \left( \frac{ T  }{  T^{*}_{\gamma} }  \right)
		  \: ,
\end{align}
where
\begin{align*}
\gamma_{0} & = \frac{2}{3} \pi^2  \, k_{\mathrm{B}}^2 \, \nu(0),
\\[3pt]
\gamma_1 & = \frac{14}{15}  \pi^2 \, k_{\mathrm{B}}^4  \,  \nu (0) ( k_0 a )^2  \big/ \varepsilon_0^* {}^2
\: ,
\end{align*}
and $\gamma_2 $ is  a constant which contains a term of the order $(k_{\mathrm{0}} a )^2 $; $T^{*}_{\gamma} = e^{- \gamma_2/ \gamma_1 }$.

This logarithmic $T^2 \ln T$ term arises mathematically from the discontinuity of $ n (\varepsilon) $ at the Fermi surface, and hence appears universally in all 3D Fermi liquids.
Physically, this term comes from interactions between two quasiparticles whose momenta are nearly equal or differ by $2 p_0$.
In this simplest model, since $\gamma_1 > 0 $, $\gamma(T)$ decreases with $T$ at low temperatures in agreement with liquid ${}^3 \mathrm{He}$.
For electrons in solids, the situation becomes much more complicated; here more than one Fermi sphere may appear, and the electron-phonon interaction also contribute a $ T^2 \ln T $ term to $\gamma(T)$.
In real solids, $\gamma_1$ is positive for $\mathrm{UPt_3}$, and negative for $\mathrm{CeAl_3}$, $\mathrm{Ce Cu_4 Ga}$, etc \cite{12,13}.


\begin{figure}[!h]
  \centering
    \includegraphics[width=\linewidth]{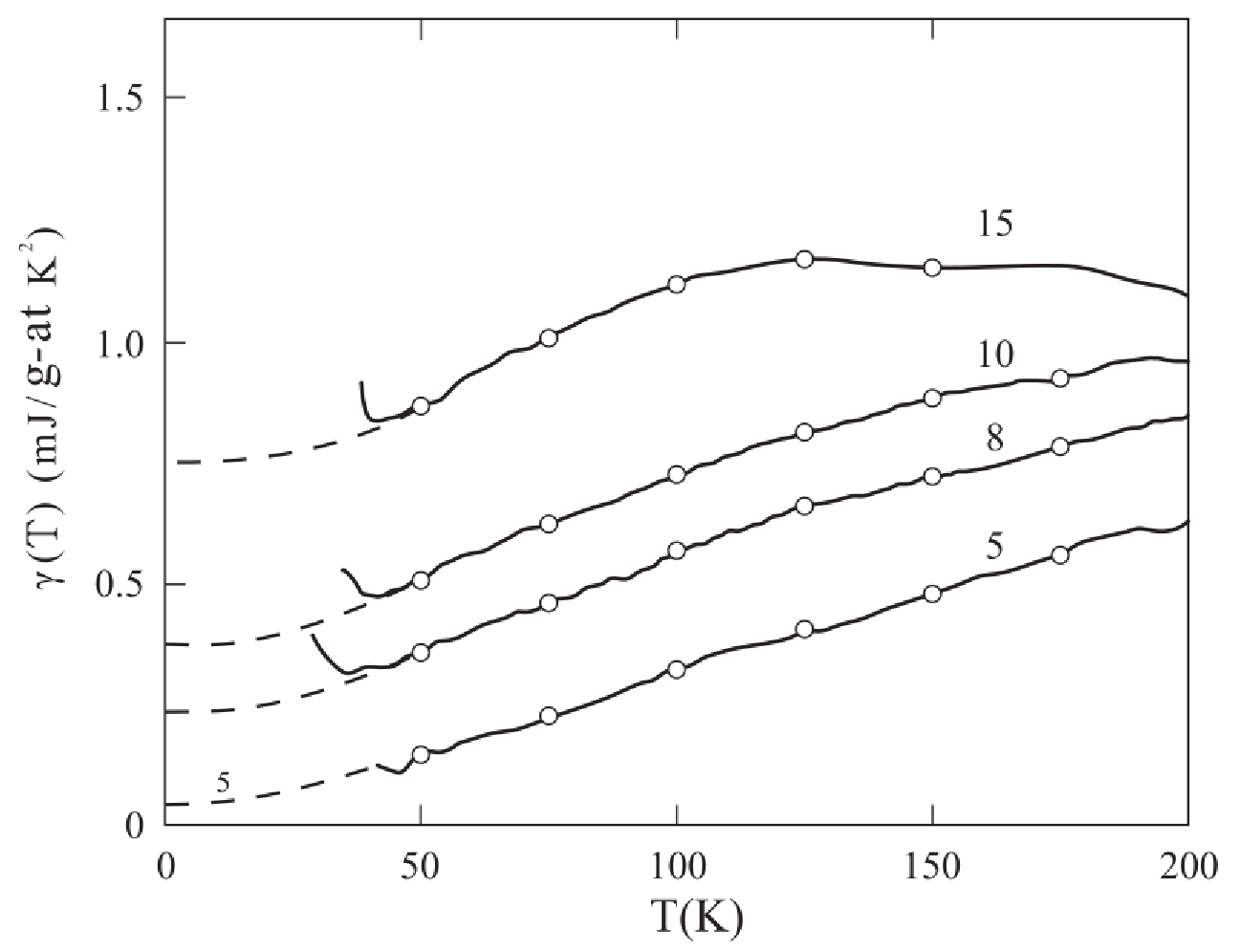} 
    \caption{
    Electronic specific heat coefficient $\gamma(T) $ vs $T$ for $\mathrm{La}_{2-x} \mathrm{Sr}_{x} \mathrm{CuO}_4 $ samples.
    Solid lines are observed data \cite{11}.
    Circles show the representative points of theoretical fit, Eq.~\eqref{2.7} in the text; dashed lines show, for reference, the extension of theoretical curves up to the lowest temperature.
    Labels show $x  \, \%  \mathrm{Sr}$.
    }   \label{fig1}
\end{figure}


For $\mathrm{La}_{2-x} \mathrm{Sr}_{x} \mathrm{CuO}_4$ compounds, $\gamma(T)$ was measured by Loram et al \cite{11};
the outline of their result is reproduced in Fig. \ref{fig1}.
Facing these data, they take notice of almost $T$-linear behaviours of $\gamma(T)$.
In fact, if we watch $\gamma (T)$ of $x=0.08$ sample, for example, $\gamma(T)$ increases almost linearly with $T$ for $50 \; \mathrm{K} \lesssim T \lesssim 100 \; \mathrm{K}$; they have concluded that $\nu(\varepsilon)$ varies as $|\varepsilon|$.
If the system were assumed to be 2-dimensional, however, the sign of the $ | \varepsilon | $ term contradicts theory \cite{14}.

This difficulty can be solved by noting that the \\ $T^2 \ln ( T/ T^{*} )$ function, which frequently appears in 3D Fermi liquids, behaves almost linearly for a certain range of $T$.
Numerically, for $15 \lesssim x \lesssim 35$,
\begin{align}
x^2 \ln \frac{ 100 }{ x } \, \fallingdotseq \, -217.6 + 43.16 x \: ,
\end{align}
within $0.6 \, \%$ scattering.

\begin{figure}[!htb]
	\centering
    \includegraphics[width=\linewidth]{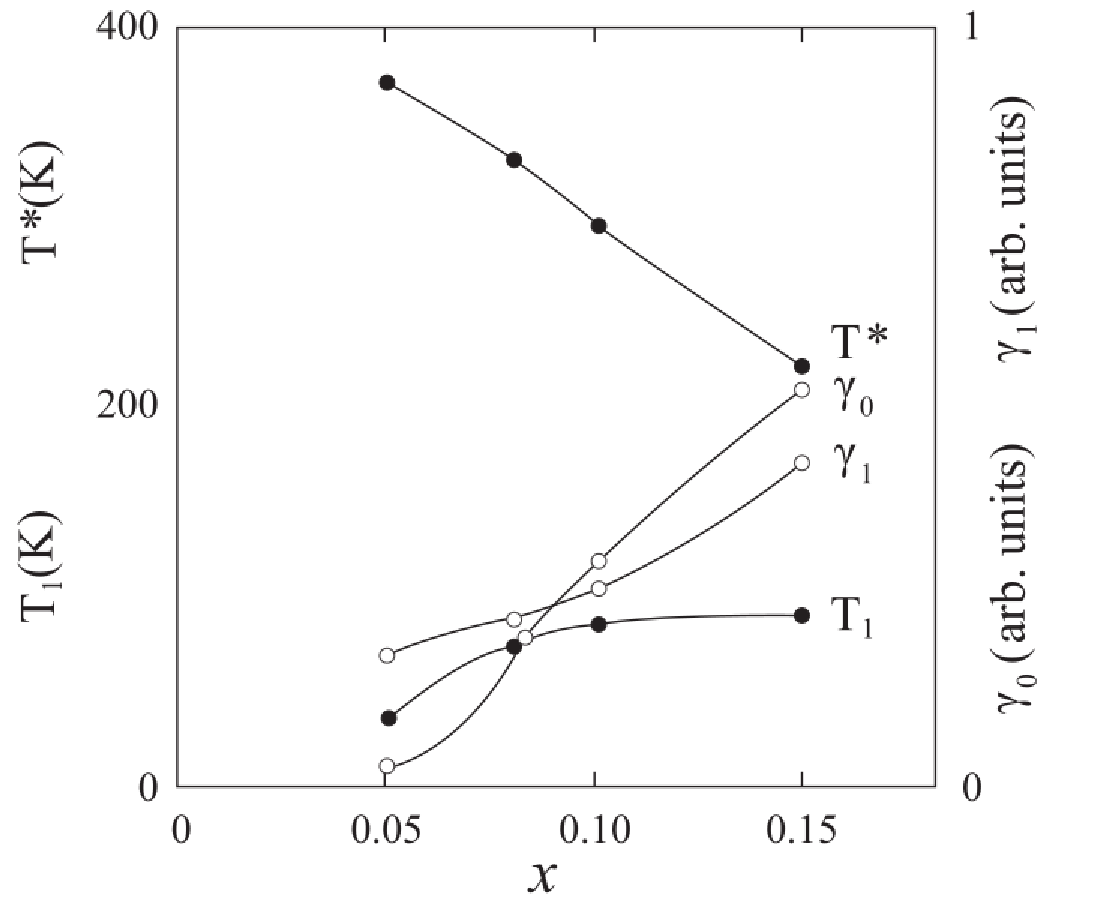} 
   	\vspace{3pt}
	\caption{
	Constants appearing in $ \gamma (T) $ of $ \mathrm{La}_{2-x} \mathrm{Sr}_x \mathrm{Cu} \mathrm{O}_4 $ samples, Eq.~\eqref{2.7}, are plotted as functions of $x$.
	Solid lines show guides to the eye.
	}
	\label{fig2}
	\end{figure}

We have analysed their data of 4 samples, $ 0.05 \leq x \leq 0.15 $, to fit the formula
\begin{subequations}
\begin{align}
\gamma(T) = \gamma_0 - \gamma_1 T^2 \ln ( T / T^{*}_{\gamma} ) \: ,
\end{align}
or
\begin{align} \label{2.7}
\gamma(T)  =  \gamma_0  \left\{  1 -  \left(  \frac{  T  }{  T_1  }  \right)^2  \ln \frac{  T  }{  T^{*}_{\gamma}  }  \right\} \: ,
\end{align}
\end{subequations}
where $\gamma_0$, $\gamma_1$, $T_1$ and $T^{*}_{\gamma}$ are constant.
As shown in Fig.~\ref{fig1}, all data are precisely fitted, within experimental scattering, to the $ T^2 \ln T $ variation. 
For $ x= 0.05 $ sample 
\begin{align}
\gamma(T)  =  0.0363  \left\{  1 -  \left(  \frac{  T  }{  40.6  }  \right)^2  \ln \frac{  T  }{  371  }  \right\}  \:\: \mathrm{mJ/mol\cdot K^2} ,
\end{align}
while, for $x= 0.15$ sample,
\begin{align}
\gamma(T)  =  0.849  \left\{  1 -  \left(  \frac{  T  }{  102  }  \right)^2  \ln \frac{  T  }{  165  }  \right\}  \:\: \mathrm{mJ/mol\cdot K^2} ,
\end{align}
In Fig.~\ref{fig2}, $x$-dependence of the constants is shown; since all curves are smoothly varying functions of $x$, the analysis is almost justified.

As the conclusion of this section, we shall state that quasiparticles in cuprates are not 2-dimensional (2D) but 3-dimensional.
According to the many body theory \cite{19''',20'''}, the leading term of the imaginary part in the self energy part $\Sigma (\varepsilon)$ for 2D fermion systems behaves as $ a \varepsilon | \varepsilon | \ln | \varepsilon| $, where $a$ is a positive constant depending on the interaction strength and number density.
Because of the dispersion relation, the corresponding real part of $ \Sigma ( \varepsilon ) $ becomes $ ( \pi / 2 ) a \varepsilon | \varepsilon | $ which gives rise to $ - ( \gamma^* / 2 )T^2 $ term to the entropy, where $ \gamma^* $ is a positive constant being proportional to $a$.
In conclusion, the specific heat coefficient is given by 
\begin{align} \label{gamma2}
\gamma_{\mathrm{I \! I}} = \gamma_0 - \gamma^* T   \:\: ; 
\end{align}
$\gamma_{\mathrm{I \! I}} $ is a linearly decreasing function of $T$ at lowest temperatures \cite{17'}.

\begin{figure}[!bh]
		\centering
	    \includegraphics[width=\linewidth]{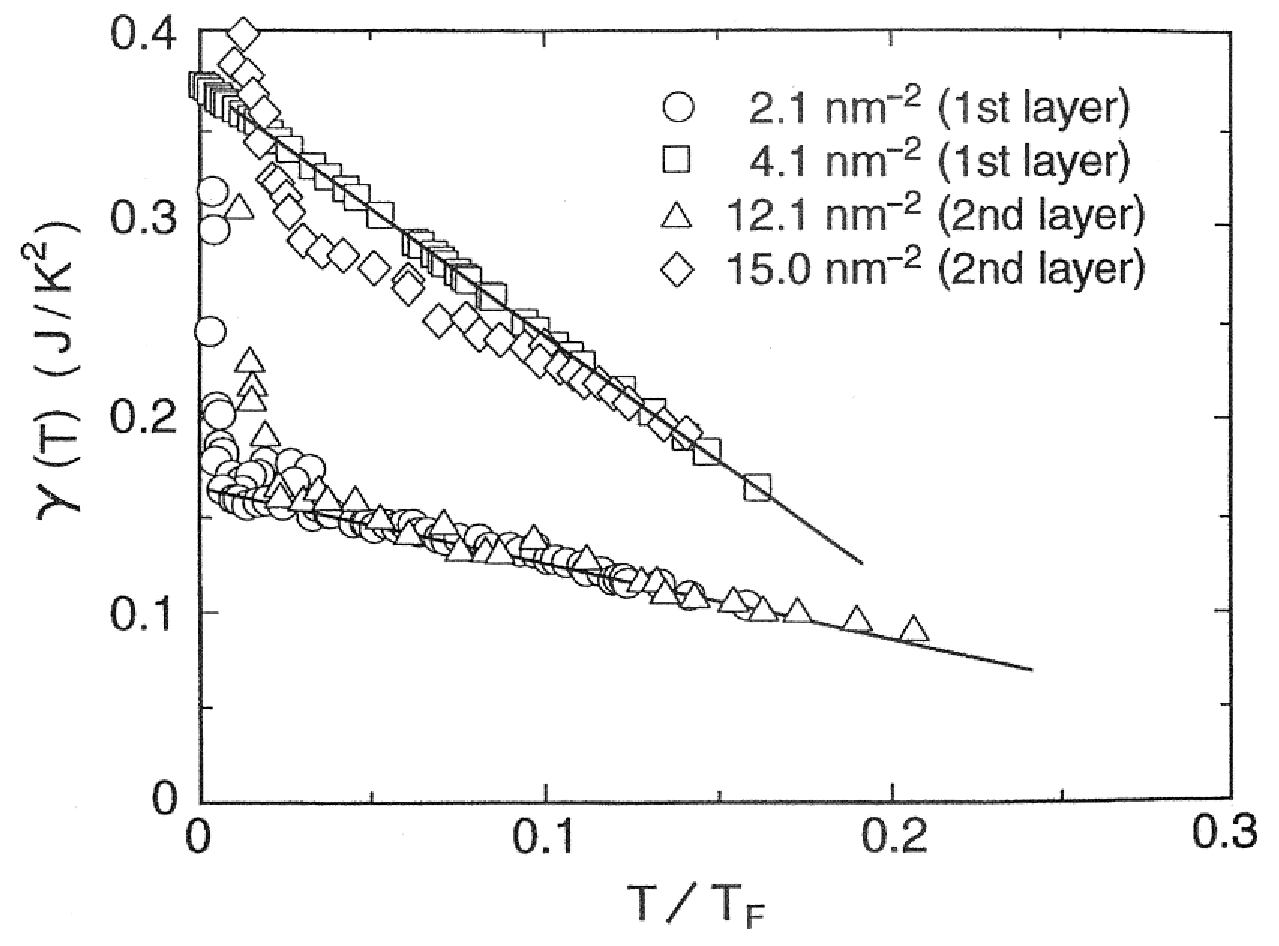}
		\caption{
		Specific heat coefficient of 2D ${}^3 \mathrm{He} $ liquids at various densities, the first and second layer films on graphite, plotted as a function of $T/T_{\mathrm{F}}$, $T_{\mathrm{F}}$ being the Fermi temperature \cite{17''}.
		}
		\label{fig3'}
	\end{figure}

In order to confirm Eq.~\eqref{gamma2} experimentally, Morishita \cite{17''} has tried to measure the heat capacity of ${}^{3} \mathrm{He}$ films.
He has measured the heat capacity of the first and second layer ${}^{3} \mathrm{He}$ films absorbed on a bare graphite surface, and found that, as shown in Fig.~\ref{fig3'}, the heat capacities divided by temperature clearly follow Eq.~\eqref{gamma2}.

It is concluded that the quasiparticles in the normal state cuprates are not 2-dimensional but 3-dimensional.


	

\begin{figure}[!htb]
	\centering
    \includegraphics[width=\linewidth]{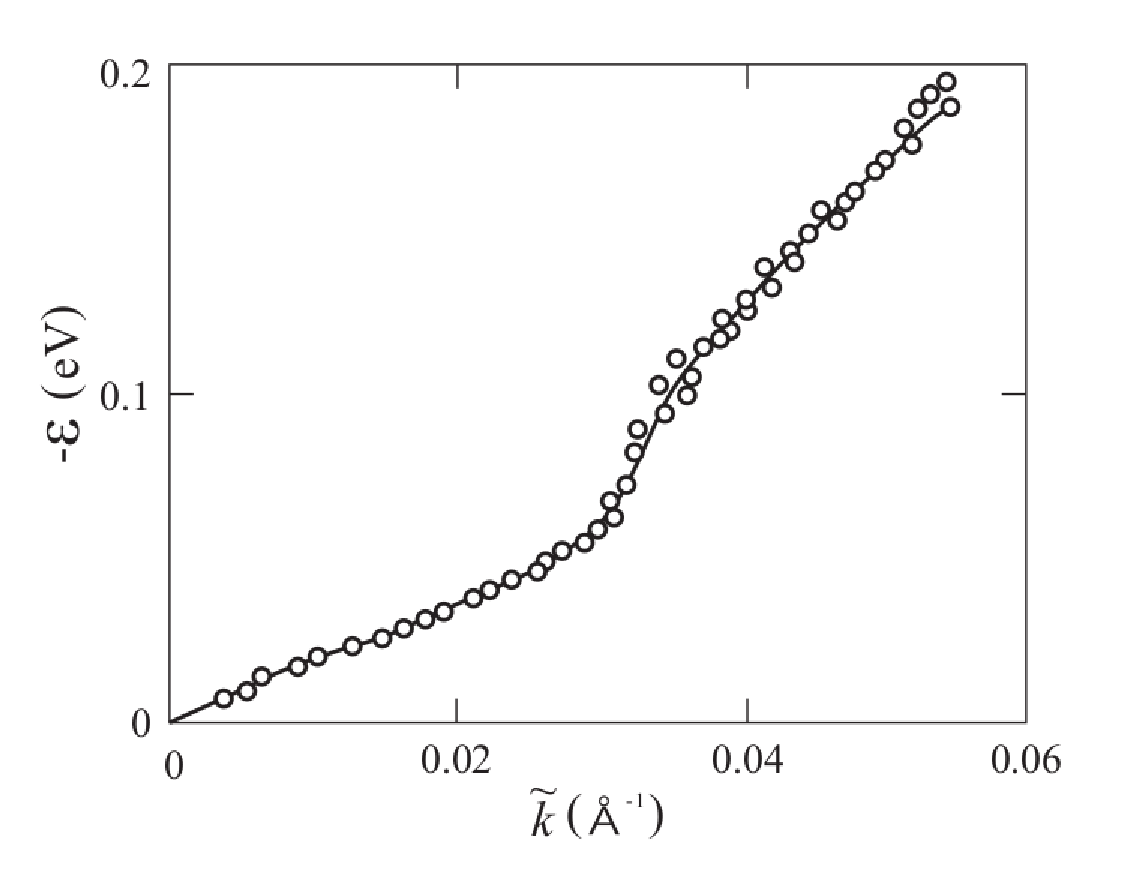} 
	\caption{
	The energy-momentum relation of the quasiparticle for $ \mathrm{La}_{2-x} \mathrm{Sr}_x \mathrm{Cu} \mathrm{O}_4 $ $ x = 0.063 $ sample.
	Circles show the experimental points \cite{15}, and solid line shows the theoretical fit, Eq.~\eqref{3.1} in the text.
	}
	\label{fig3}
\end{figure}

\section{Kink Phenomena of LSCO}

The quasiparticle behaviours in the interacting electron system are usually described by the energy-momentum relation, $ \varepsilon(p) = \varepsilon(\hbar k) $, and the energy dependence of the self-energy part, $\Sigma ( \varepsilon)$.
Experimentally, these problems have been very extensively examined for $\mathrm{LSCO}$ by Zhou and his group \cite{15} through the angle-resolved-photoemission spectroscopy.

Concerning the $\varepsilon$ vs $\tilde{k} $ relation, they have found that an abrupt slope change (kink) appears in $ x = 0.03$, $0.063$ and $0.07$ samples, where $\tilde{k}  = k- k_0$, $k_0$ being the Fermi wave number.
This anomalous behaviour may be related to the $ \tilde{p}^3 \ln | \tilde{p} | $, or $ \tilde{k}^3 \ln | \tilde{k} | $, dependence of $\varepsilon (p) $, Eq~\eqref{1.2}.
We shall try to fit the data of $ x = 0.063 $ sample which is  reproduced in Fig. \ref{fig3}.
We choose 12 points for $ 0\leq \tilde{k} \leq 0.055\, \mathrm{\AA}^{-1} $, and carefully read the corresponding values of $\varepsilon$.
These points can be fitted by the relation,
\begin{align} \label{3.1}
- \varepsilon
 \, =\,
 4.588 \, \tilde{k}  \,
 \Big\{
 		1 - 112.6 \, \tilde{k}  -  2042.6  \, \tilde{k}^2  \ln ( 6.954 \tilde{k} )
 \Big\} \; \mathrm{eV} \: .
\end{align}
Here four figures are needed to reproduce accurately the experimental data.
Eq.~\eqref{3.1} is plotted in Fig.~\ref{fig3}.

We should comment the accuracy of Eq.~\eqref{3.1};
the root-mean-square (rms) error of this equation to the original curve is $2.6 \, \%$.
If the curve is expressed, without the logarithmic term $\tilde{k}^3 \ln \tilde{k}$, by power series up to $\tilde{k}^3 $ or $\tilde{k}^4$, the corresponding error is $7.2 \, \%$ or $3.3 \, \%$.
Thus the existence of $\tilde{k}^3 \ln \tilde{k}$  term seems to be essential for explaining this kink phenomena.

As seen in Fig.~\ref{fig3}, the kink of quasiparticle energy appears for the wave number of about $0.03 \; \mbox{\AA}^{-1}$.
As will be mentioned later for $\Sigma (\varepsilon)$, this value of the kink has no special meaning.

Next, we shall discuss the kink of the self energy part, $\mathrm{Re} \Sigma (\varepsilon)$, which appears at about $60 \, \mathrm{meV} $ in $\mathrm{LSCO}$.
The general theory for the self energy part of 3D fermion systems was discussed comprehensively by Migdal \cite{7}.
Examining the lowest order diagrams for $\Sigma$, he showed that $\mathrm{Im} \Sigma $ takes the form
\begin{align} \label{3.2}
\mathrm{Im} \, \Sigma (\varepsilon)  =  \gamma \varepsilon | \varepsilon | + \alpha \varepsilon^3  + \alpha' \varepsilon^2 | \varepsilon | \: ,
\end{align}
where $\gamma$, $\alpha$ and $\alpha ' $ are constants.
Here $ \Sigma ( \varepsilon ) $ is not an analytic function of $\varepsilon$;
the analytic function in the upper half plane, $\sigma ( \varepsilon )$, is given by
\begin{align} \label{3.3}
\mathrm{Im} \, \sigma( \varepsilon)  =  \gamma \varepsilon^2 + \alpha \varepsilon^2 | \varepsilon | + \alpha ' \varepsilon^3 \: ,
\end{align}
which states that, because of the sharp Fermi surface, $ \mathrm{Im} \, \sigma $ has a discontinuity equal to $2 \alpha \varepsilon^3$ when the sign of $\varepsilon$ changes.
Using the dispersion relation for $ \sigma ( \varepsilon ) $, he drived
\begin{align} \label{3.4}
\Sigma ( \varepsilon ) = \Sigma_0 ( \varepsilon) + ( 2/ \pi ) \alpha \varepsilon^3 \ln ( | \varepsilon | / \mu ) \: ,
\end{align}
where $\mu$ is the Fermi energy (chemical potential) , $ \mathrm{Re} \, \Sigma_0 (\varepsilon )$ is expressed in a power series of $\varepsilon$, and $\mathrm{Im} \, \Sigma_0 (\varepsilon)$ is given by Eq.~\eqref{3.2}.
The $ \varepsilon^3 \ln |\varepsilon |$ term in Eq.~\eqref{3.4} confirms the result of Galitsky, Eq.~\eqref{1.2}.
It should be noted that the logarithmic $ \alpha \varepsilon^3 \ln ( |\varepsilon| / \mu ) $ term arises exclusively from the presence of Fermi surface in 3D systems.

\begin{figure}[b]
	\centering
    \includegraphics[width=\linewidth]{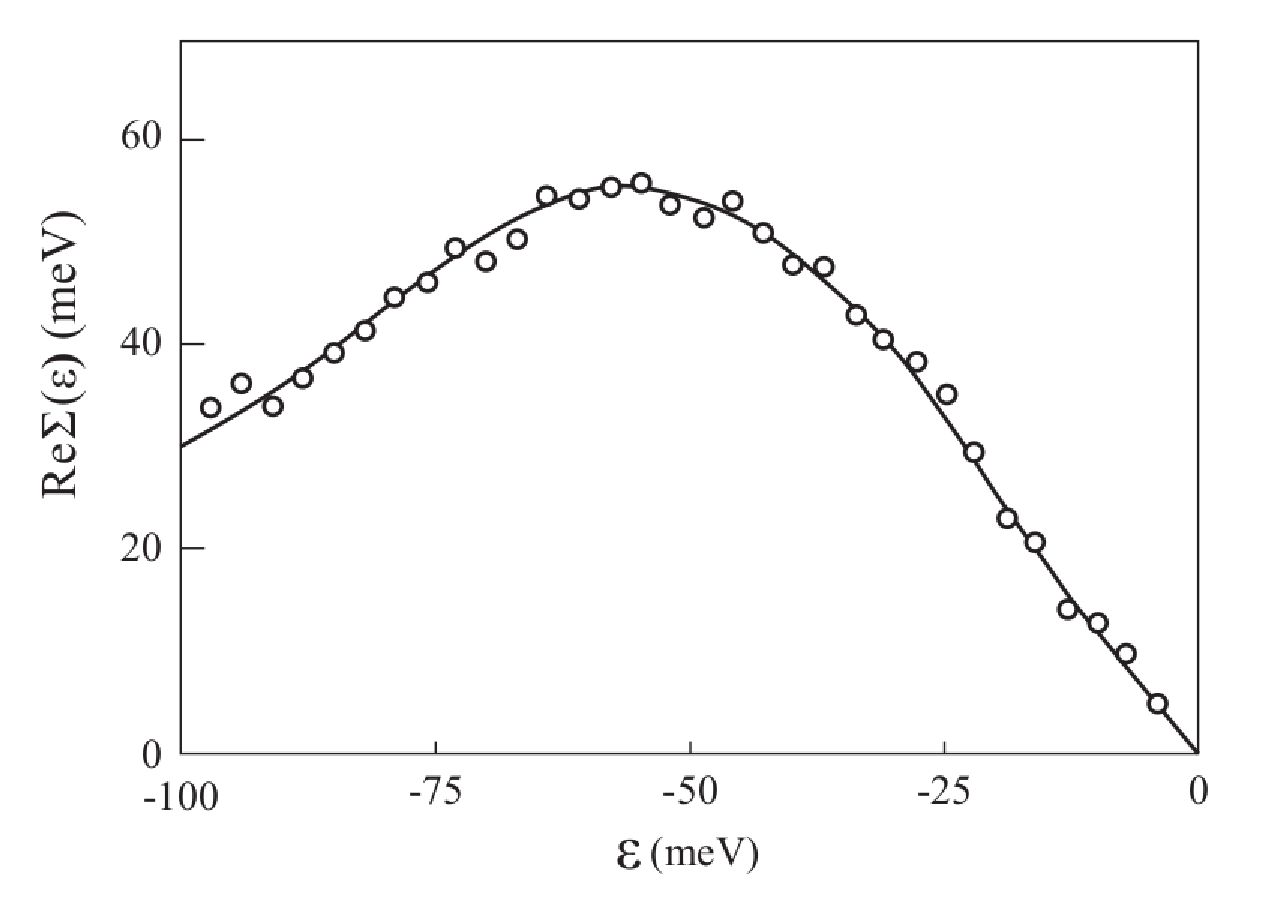}
	\caption{
		$\mathrm{Re} \, \Sigma ( \varepsilon )$ vs $\varepsilon$ relation of the quasiparticle for $ \mathrm{La}_{2-x} \mathrm{Sr}_x \mathrm{Cu} \mathrm{O}_4 $ $ x= 0.063 $ sample.
		Circles show the experimental points \cite{22}, and solid line shows the theoretical fit, Eq.~\eqref{3.5} in the text.
			}
	\label{fig4}
\end{figure}

Here we shall take $ x = 0.063$ sample again for analysing the $\varepsilon$ vs. $\mathrm{Re} \, \Sigma (\varepsilon)$ relation;
their experimental data are reproduced in Fig. \ref{fig4}.
We read ten representative points in the figure for the range $0 \leq | \varepsilon | \leq 100 \,\mathrm{meV}$;
these points are shown to obey Eq.~\eqref{3.4} precisely through the formula
\begin{align}
\mathrm{Re} \, \Sigma ( \varepsilon )
&= - 0.7357 \, \varepsilon
	\Big\{
	1 - \frac{ \varepsilon }{  11. 73  }
	\notag 
	\\
& \qquad + \left( \frac{  \varepsilon }{  33.883  } \right)^2
			\ln \frac{  | \, \varepsilon \, |  }{  284. 89  }
	\Big\} \; \mathrm{meV} \: .
	 \label{3.5}
\end{align}
Here, for $ \varepsilon^3 \ln |  \, \varepsilon \, | $ term, five figures are needed to reproduce the experimental data accurately.
Equation \eqref{3.5} is plotted in Fig.~\ref{fig4} by the solid line;
$\mathrm{Re} \, \Sigma ( \varepsilon )$ shows the maximum value $54.8 \; \mathrm{meV}$ at $\varepsilon = - 55.2 \; \mathrm{meV}$.

The accuracy of Eq.~\eqref{3.5} can be judged by examining the rms error.
If the curve is expressed, without the $\varepsilon^3 \ln | \, \varepsilon \, |$ term, by power series up to $\varepsilon^3$, or $\varepsilon^4$, the corresponding rms error is $7.82 \, \%$, or $ 0. 76 \, \% $, while that for Eq.~\eqref{3.5} is $0.59 \, \%$.
Thus we conclude that the presence of the $\varepsilon^3 \ln | \, \varepsilon \, |$ term is essential for explaining the $\varepsilon$ vs $\mathrm{Re} \, \Sigma ( \varepsilon )$ relation.

We have to note that, contrary to the general belief, the energy of the kink ($55.2 \, \mathrm{MeV}$) has no special meaning;
if we assume that the $ \varepsilon^3 \ln |\varepsilon | $ term is related to a collective or coherent motion with the characteristic energy $\sigma$, then we may have a term $ a \varepsilon^3 \ln ( |\varepsilon| /\sigma ) $, to which we have to add ordinary $ b \varepsilon^3 $ term from Eq.~\eqref{3.4}, where $a$ and $b$ are constants;
here $a$ arises exclusively from the interactions, while $b$ is the sum of an interaction-free (band) term and an interaction-dependent term. 
Since 
\[
a \varepsilon^3 \ln (|\varepsilon | / \sigma) + b \varepsilon^3  =  a \varepsilon^3  \ln ( |\varepsilon | / \sigma e^{-b/a} ) \:\: , 
\]
the kink energy is a fraction of $ \sigma e^{-b/a}  $, which is usually quite different from $\sigma$, since $e^{-b/a}$ ranges from $0$ to $\infty$, depending on the interaction strength.

Concerning the nature of $\mathrm{Re} \, \Sigma ( \varepsilon )$, Zhou et al have asserted that, when they plot the second derivative of $\mathrm{Re} \, \Sigma (\varepsilon)$, there appear 3 or 4 peaks corresponding to phonon's characteristic energies.
We should examine this on the basis of electrons interacting with the phonon of the energy $\hbar \omega$.
Migdal \cite{16} has shown that $ \mathrm{Re} \, \Sigma (\varepsilon) $ due to this interaction is given by
\begin{align} \label{3.6}
\delta \mathrm{Re} \, \Sigma ( \varepsilon)
	=
	\frac{1}{6}  \,  \frac{ \lambda }{  1 - 2  \lambda  }   \,
		\frac{  \varepsilon^3  }{  \:  u^2  p_0^2  \:  }
			\ln \frac{  | \, \varepsilon \, |  }{  \:  \hbar \omega  \:  } \: ,
\end{align}
where $\lambda$ denotes the interaction strength, and $u$ is the sound velocity.
If electrons are interacting with $n$ phonon modes, $\varepsilon_i = \hbar \omega_ i \: ( i = 1,2, \cdots, n)$, $\mathrm{Re} \, \Sigma ( \varepsilon )$ is given, in addition to the $\varepsilon$ and $\varepsilon^2$ terms, by
\begin{align}\label{3.7}
\mathrm{Re} \, \Sigma ( \varepsilon)
 =
 a_0 \varepsilon^3 \ln \frac{  | \, \varepsilon \, |  }{  \varepsilon_0  }
 +  \underset{i=1}{\overset{n}{\Sigma}}   a_i  \varepsilon^3  \ln \frac{  | \, \varepsilon \, |  }{  \varepsilon_i }
 \: ,
\end{align}
where the first term comes from the electron-electron interaction, and
\begin{align*}
a_i = \frac{1}{6}  \, \frac{ \lambda }{  1 -  2 \lambda   }  \,
			\frac{  1  }{ \:  u_i^2  p_0^2 \: } \: \: , \quad i =1,2,\cdots, n \; .
\end{align*}
Equation \eqref{3.7} can be summed up into one term,
\begin{align} \label{3.8}
\mathrm{Re} \, \Sigma ( \varepsilon ) = \left( \underset{i=0}{\overset{n}{\Sigma}} a_i \right) \varepsilon^3 \ln \frac{  | \, \varepsilon \, |  }{  \varepsilon^{*}  } \: ,
\end{align}
where
\begin{align*}
\varepsilon^{*}  =  \exp \left\{   \sum_{i=0}^n  a_i  \ln  \varepsilon_i  \bigg/  \sum_{i=0}^n  a_i   \right\} \: .
\end{align*}
The peak of the second derivative of $\mathrm{Re} \, \Sigma ( \varepsilon )$ is determined by $\mathrm{d}^3 \mathrm{Re} \, \Sigma (\varepsilon) / \mathrm{d} \varepsilon^3  =  0  $  which gives  $\varepsilon_{\mathrm{peak}} = e^{  - 11/ 6 }  \varepsilon^{*}  \fallingdotseq 0.16 \, \varepsilon^{*}$.
To the contrary of Zhou et al.'s assertion, the peak is only one and never appears at each $ \hbar \omega_i $.
Valla \cite{17} has already stated that the fine structure of $\mathrm{Re} \, \Sigma (\varepsilon)$ cannot be real and is probably noise related.

\section{Magnetic Susceptibility, Nuclear Spin-Lattice Relaxation Time and Knight Shift}

A general and exact formula for the magnetic susceptibility at finite temperatures, was given, on the basis of Landau's spirit, by Usui \cite{18};
\begin{align} \label{4.1}
\chi  =  2 \beta  \mu_{\mathrm{B}}^2  \sum_i \sum_j \left(  j  \left| \frac{ 1 }{   1  +  2 \beta \psi  } \right|  i  \right) \tilde{\nu}_i \tilde{\nu}_j  \: ,
\end{align}
where $\mu_{\mathrm{B}} $ is the magnetic moment of one particle, $i$ and $j$ refer to the single quasiparticle state with energy $\varepsilon_i$ and $\varepsilon_j$, their distribution function being $ n_i  =  \big( e^{\beta \varepsilon_i }  +  1 \big)^{-1} $, $ \tilde{\nu}_i  = [ n_i ( 1- n_i )]^{1/2} $.
$( j | ( 1 + 2 \beta \psi )^{-1} | i) $ is the $(j,i)$ element of the inverse matrix $(1 + 2 \beta \psi )^{-1}$; $\psi_{ij} = \tilde{\nu}_i g_{ij} \tilde{\nu}_j$, $g=  (1/2)  ( f_{\uparrow \uparrow}  -  f_{\uparrow \downarrow}) $ being the spin antisymmetric part of the Landau $f$-function.
When the electron (or hole) density is relatively low, the $g$ function can be described, as a function of momenta $\bm{p}$ and $\bm{p} ' $, by the sum of ladder type diagrams for quasiparticles interacting with the screened Coulomb interaction which is specified by the s-wave scattering length $a$;
\begin{align} \label{4.2}
g ( \bm{p}, \bm{p}' )
	&=
		- \frac{  2 \pi \hbar^2 a  }{  m V  }  +  \frac{  4  \hbar  a^2  }{  m \pi V  } \int  d \bm{p}_1  n_0  (  \bm{p}_1  )
\notag \\
	& \: \quad  \times
	 \int d\bm{p}_2 \frac{  \delta(  \bm{p} + \bm{p}' - \bm{p}_1  - \bm{p}_2  )   }{  p^2  + p'{}^2  - p_1^2  - p_2^2  }
	\: ,
\end{align}
where $V$ is the volume of the system, $n_0 ( \bm{p})$ is the Fermi distribution function at absolute zero and $ p= |\bm{p}| $.
After performing integration in Eq.~\eqref{4.2} and taking the average over the angle between $\bm{p}$ and $\bm{p}'$, we obtain $g ( p, p')$.
When the quasiparticles are situated near the Fermi surface, $g(p,p')$ contains terms like $(p-p_0)^2 \ln | p-p_0 |$ or $(p'-p_0)^2 \ln | p' - p_0 |$.
It should be noted that these logarithmic terms arise exclusively from short-wavelength charge and spin-density fluctuations whose wave number is about $2 k_{\mathrm{0}} \: ( k_{\mathrm{0}} = p_0 / \hbar )$.
When inserted into Eq.~\eqref{4.1} these terms produce the $ T^2 \ln T $ dependence of $\chi(T)$;
\begin{align} \label{4.3}
\chi (T)
	&=  \chi (0)  - bT^2 \ln (T/T^*) + cT^2 + \mathrm{O} ( T^4 \ln T, T^4 ) \:
\notag \\
	&=  \chi (0) - bT^2 \ln ( T/ T^*_{\chi} ) + \mathrm{O} ( T^4 \ln T, T^4 ) \: ,
\end{align}
where $\chi(0)$, $T^*$ and $c$ are constant, $ T^{*}_{\chi}  =  T^{*} e^{c/b} $ and $b$ is given by
\begin{align} \label{4.4}
b  =  \frac{1}{3}  ( k_{\mathrm{0}} a )^2 \left( \frac{ k_{\mathrm{B}} }{  \varepsilon^{*}_{0}  } \right)^2  \frac{  \chi (0)^2  }{  \chi_0^0  } \: ,
\end{align}
Here $\chi_0^0 = 3 m n_0 \mu_{\mathrm{B}}^2 / p_0^2$, $n_0$ being the number density of the system.
Since $ b > 0 $, the susceptibility increases with $T$ and exhibits universally a maximum at temperature $ T_{\mathrm{max}} =  T^{*}_{\chi} / \sqrt{e} $.
Thus we predict the phenomenon of the susceptibility maximum which should appear in all paramagnetic Fermi liquids.

Here, $T_{\mathrm{max}}$ has no special physical meaning; this is very important; since, in the literature, the importance of $T_{\mathrm{max}}$ is emphasized in relation to the spin gap phenomena.
In $T^*_{\chi} = T^* e^{c/b}$, $b$ arises from the particle interaction, while $c$ is a sum of the non-interacting (band) term and interaction-dependent term;
thus $e^{c/b}$ varies from $0$ to infinity depending on the interaction strength;
regarding $d$-metals, $T^{*}_{\chi}$ for $\mathrm{Pd}$ is $132 \,\mathrm{K}$, while that for $\mathrm{Ir}$ is $ 11,300 \, \mathrm{K} $ \cite{19}.
When  the system  becomes nearly ferromagnetic, $\chi (0)$ is enhanced by the Stoner (or Landau) enhancement factor $L$, $T^{*}$ being reduced by $1/L$;
since the dominant term of $b$ and $c$ is known to be proportional to $L^2$, $c^{c/b}$ keeps constant.
Thus we have
\begin{align} \label{4.5}
\chi (0) T_{\mathrm{max}} \approx \text{const.}
\end{align}
%


\begin{figure}[h]
	\centering
		\includegraphics[width=\linewidth]{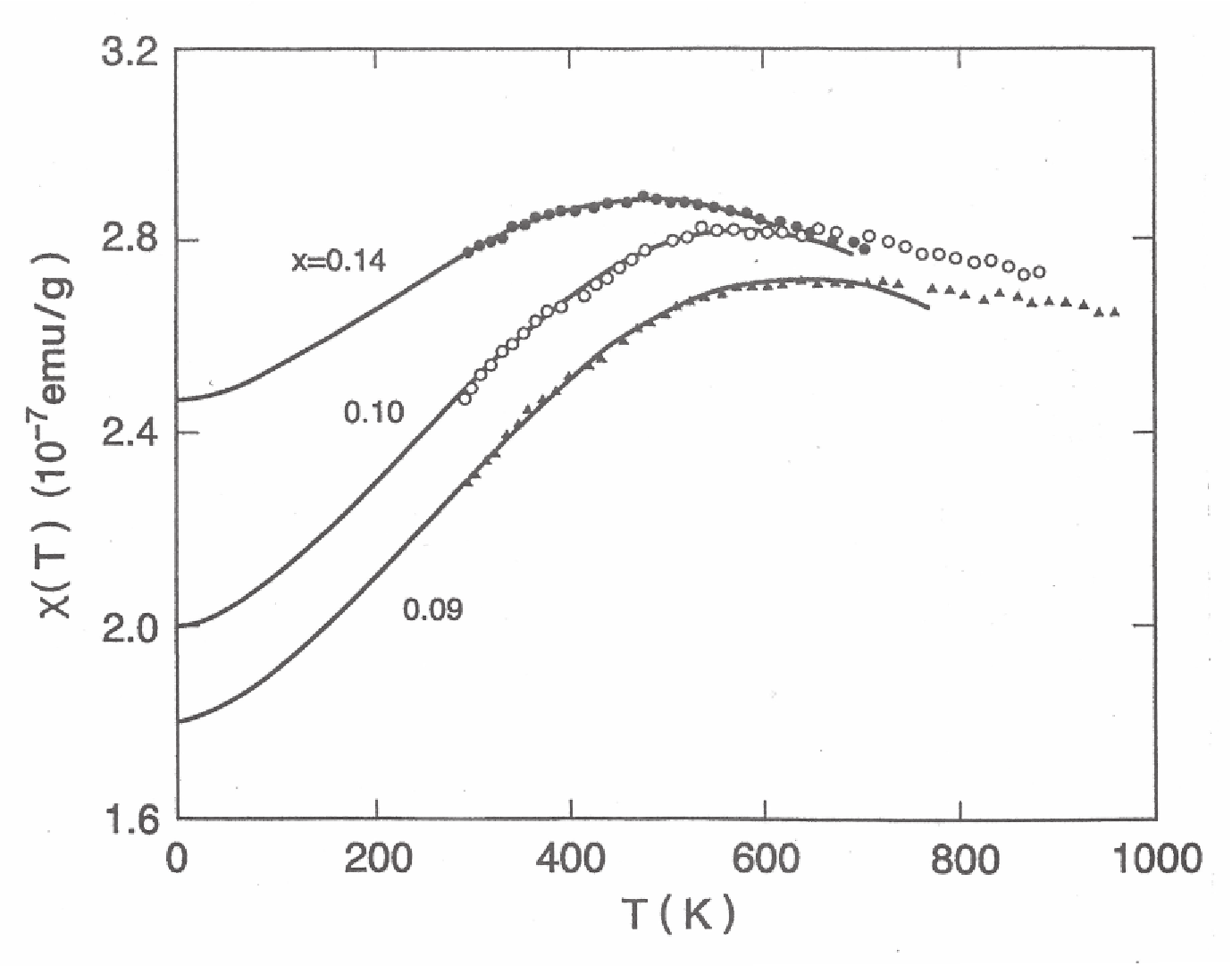}
	\caption{
		Magnetic susceptibility $\chi(T)$ vs $T$ of $ \mathrm{La}_{2-x} \mathrm{Sr}_x \mathrm{Cu} \mathrm{O}_4 $ samples.
		Symbols; experimental data \cite{20}.
		Solid lines; theoretical curves, Eq.~\eqref{4.6} in the text.
			}
	\label{fig5}
\end{figure}


The magnetic susceptibility of $\mathrm{LSCO}$  was measured by many groups; among them we adopt to analyse the experimental data of Yoshizaki et al. \cite{20}, since the scattering of the data is small;
the main part of the data is reproduced in Fig. \ref{fig5}.
Here we express $\chi (T)$ in the form
\begin{align} \label{4.6}
\chi (T) = \chi (0 ) \left\{ 1 - \left( \frac{T}{T_0} \right)^2 \ln \frac{T}{T^{*}_{\chi}} \right\} \: ,
\end{align}
where $T_0$ and $T^{*}_{\chi} $ are constant.
As shown in the figure, the observed data can be precisely fitted by Eq.~\eqref{4.6}.
The result of the analysis is listed in Table \ref{tab1}.
It is to be noted that, although the system is not nearly ferromagnetic, the relation, $\chi(0) T_{\mathrm{max}} \approx \text{const.} $, is satisfied.

\begin{table*}[!bh]
\centering
\caption{Constants concerning the magnetic susceptibility of LSCO, Eq.~\eqref{4.6}. $\gamma(0)$ values are due to Loram et. al \cite{11}.}
\label{tab1}
\renewcommand{\arraystretch}{1.3}
\begin{tabular}{ccccccc} \hline
 $x$ &
 \begin{tabular}{c} $\chi(0)$ \\ {\small $10^{-5} \mathrm{emu/mol}$}  \end{tabular} &
 \begin{tabular}{c}  $\gamma(0)$  \\  {\small $\mathrm{mJ / mol}$ }  \end{tabular} &
 \begin{tabular}{c}  $ T_0 $  \\  {\small $\mathrm{ K }$ }  \end{tabular} &
 \begin{tabular}{c}  $ T_{\mathrm{max}} $  \\ {\small $\mathrm{ K } $}  \end{tabular} &
 \begin{tabular}{c}  $ \chi (0) T_{\mathrm{max}} $  \\  {\small $\mathrm{ arb. units }$}  \end{tabular} &
 $ F_0^a $ \\ \hline
 0.09 & 7.21 & 0.308 & 631 & 640 & 4.61 & $- 0.415$  \\
 0.10 & 7.98 & 0.377 & 641 & 585 & 4.67 & $- 0.353$  \\
 0.14 & 9.82 & 0.745 & 834 & 485 & 4.76 & $  0 .043$  \\ \hline
  \end{tabular}
\renewcommand{\arraystretch}{1}
\normalsize
\end{table*}

The magnetic state of the system can be specified by evaluating the Landau parameter $F_0^a$ through the formula
\begin{align} \label{4.7}
\frac{ \chi (0 )}{ \gamma ( 0 ) } = \frac{ 3 }{ \pi^2} \frac{ \mu_{\mathrm{B}}^2 }{ k_{\mathrm{B}}^2 } \frac{ 1 }{ 1 + F_0^a } \: .
\end{align}
The $\gamma (0)$ and $F_0^a$ values of LSCO are listed in Table \ref{tab1}.
For $x = 0.09$ sample, $\chi (0)$ is enhanced by the factor $ 1/ ( 1-0.415)  \fallingdotseq 1.71 $ compared with the value determined by $\gamma(0)$.
Thus, LSCO is neither nearly antiferromagnetic, nor nearly ferromagnetic, but enhanced paramagnetic.
In literatures, the role of antiferromagnetic fluctuations is greatly emphasized for high-$T_{\mathrm{C}}$ cuprates.
%
This view should be amended;
the role of charge fluctuations and that of spin fluctuations are equally important.


\begin{figure}[!t]
	\centering
    \includegraphics[width=\linewidth]{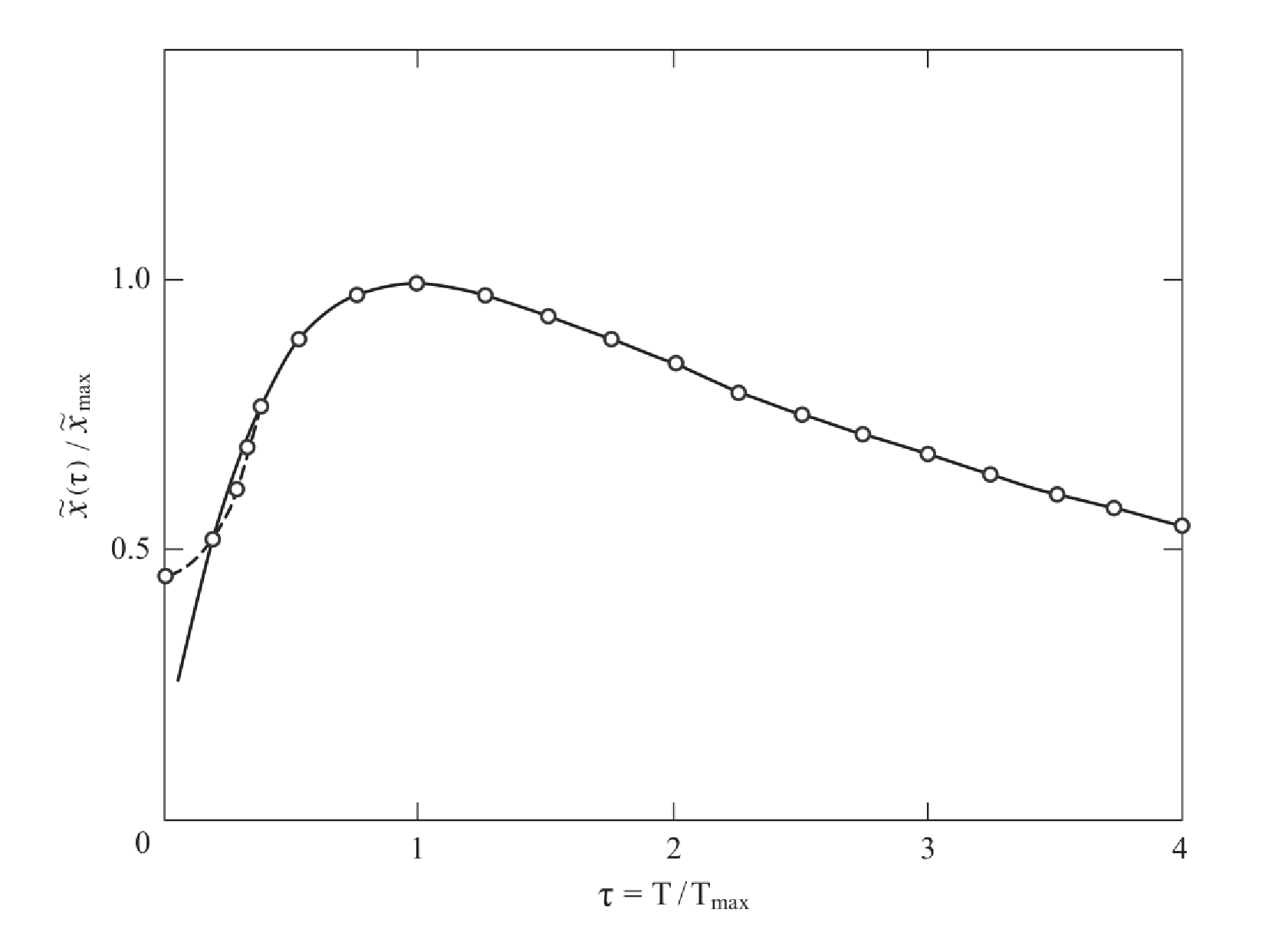}
	\caption{
	Universal curve $\tilde{\chi} (\tau) / \tilde{\chi}_{\mathrm{max}} $ vs $\tau$ of $ \mathrm{La}_{2-x} \mathrm{Sr}_x \mathrm{Cu} \mathrm{O}_4 $ samples determined by Nakano et al. \cite{22} is plotted by the solid line.
	Circles and dashed line are theoretical fit, Eq.~\eqref{4.8} in the text.
	For an explanation of the discrepancy between theory and experiment at lowest temperatures, see the text.
	}
	\label{fig6}
\end{figure}


Concerning the temperature dependence of the susceptibility for LSCO, Johnston \cite{21} has found a remarkable law.
Chose $\chi_0$ for a fitting parameter and define $\tilde{\chi} (T) = \chi(T) - \chi_0 $;
$\tilde{\chi} (T)$ has a maximum at $ T = T_{\mathrm{max}} ; \, \tilde{\chi} ( T_{\mathrm{max}} )\equiv  \tilde{ \chi }_{\mathrm{max}} $.
When $\tilde{\chi} (T) / \tilde{\chi}_{\mathrm{max}}$ curves for various $x$ samples are plotted as a function of $\tau = T / T_{\mathrm{max}}$, they fall on a universal curve.
Nakano et al. \cite{22} have applied this law to 11 samples of different $x$, $0.10 \leq x \leq 0.26$, and found a universal curve shown in Fig. \ref{fig6}.
We analyse this curve on the basis of the Fermi liquid picture;
we find that the curve can be fitted by $ \tau^2 \ln \tau $ term up to $ \tau \lesssim 1.4$, by $ \tau^2 \ln \tau + \tau^4 \ln \tau $ up to $ \tau \lesssim 2 $, and finally we obtain
\begin{align} \label{4.8}
\tilde{\chi} (\tau) / \tilde{\chi}_{\mathrm{max}}
	&=
	0.450 \, \bigg\{ \,
		1 - \Big( \frac{ \tau }{ 0.401 } \Big)^2  \ln  \frac{ \tau  }{  0.754  }
\notag \\
	& \quad
		-  \Big(  \frac{  \tau  }{  0.791  }  \Big)^4  \ln  \frac{  \tau  }{  2.98  }
		-  \Big(  \frac{  \tau  }{  1.544  }  \Big)^6  \ln  \frac{  \tau  }{  13.1  }
	\, \bigg\} \: ,
\end{align}
which fits the curve for the range $ 0.4 \lesssim \tau \lesssim 4 $ very precisely (rms error $0.23\, \%$).
For $\tau \lesssim 0.4 $, the universal curve seems to behave $T$-linearly in contradiction to our prediction.
For an interacting or non-interacting fermion system, however, $\chi (T)$ does not have the $T$-linear term at lowest temperatures.
In fact, our prediction agrees with original Johnston's curve \cite{21}.
Here, to be remarkable in Eq.~\eqref{4.8}, temperatures which characterize the magnitude of logarithmic terms lie in the ratio $0.401:0.791:1.544 \approx 1:2:4$, while those showing the range of logarithmic terms are $0.754:2.98:13.1 \approx 1 : 2^2 : 4^2$; it is not clear, however, whether these rules have physical origin or not.

We shall describe the $T$-dependence of the nuclear spin-lattice relaxation time $1/ T_1 T$.
Yasuoka et al. \cite{12'} have performed extensive experimental studies to find a peak structure in the $T$-dependence of $1/T_1 T$ for YBCO compounds; 
because of a rapid decrease of $1/T_1 T$ below $T_{\rm max}$, they attribute it to the spin-gap phenomena.

Here we describe this on an alternative viewpoint.
As mentioned in the Introduction, according to the experimental observations by Alloul et al. \cite{13'''} and Horvatic et al. \cite{14'''}, the $T$-dependence of relaxation time, $ 1/T_1 T $, has been shown to follow that of the Knight shift $ K (T) $ or the susceptibility $\chi(T)$; 
$ 1/ T_1 T $ varies as $ T^2 \ln T $ and universally exhibits a maximum.
As the typical examples, we may list the $ 1 / T_1 T $ for $ \mathrm{Cu} (2) $ sites in $ \mathrm{YBa}_2 \mathrm{Cu}_{4} \mathrm{O}_{8} $ \cite{12'}, and for $ \mathrm{Cu}^{63} $ sites in underdoped $ \mathrm{YBa}_2 \mathrm{Cu}_{3} \mathrm{O}_{6.64} $ \cite{23}; the analysed results are 
\begin{subequations}
\begin{align} \label{26}
1/T_1 T  =  3.295 \left\{ 1 - \Big( \frac{T}{110} \Big)^2 \ln \frac{T}{266} \right\} \; \mathrm{sec^{-1} \! \cdot  K^{-1} } \:\: .
\end{align}
for the former, and 
\begin{align} \label{26b}
1/ T_1 T = 0.3728 \Big\{ & 1 - \Big( \frac{T}{16.8} \Big)^2  \ln \frac{T}{ 126}
\notag  \\[5pt]
& - \Big( \frac{T}{71.8} \Big)^4 \ln \frac{T}{769} \Big\} 
\;
\mathrm{sec^{-1} \cdot K^{-1}} 
\end{align}
\end{subequations}
for the latter.
As shown in Fig. \ref{fig8'}, these relations precisely reproduce the experimental data from $T_{\mathrm{C}} \, (81 \mathrm{K})$ to $200 \; \mathrm{K}$, and $75 \; \mathrm{K}$ to $300 \; \mathrm{K}$.
The $1/T_1 T$ maxima appear at $ 266/ \sqrt{e} = 161 \; \mathrm{K} $, and $140 \; \mathrm{K}$.

\begin{figure}[!tb]
\centering  \includegraphics[width=\linewidth]{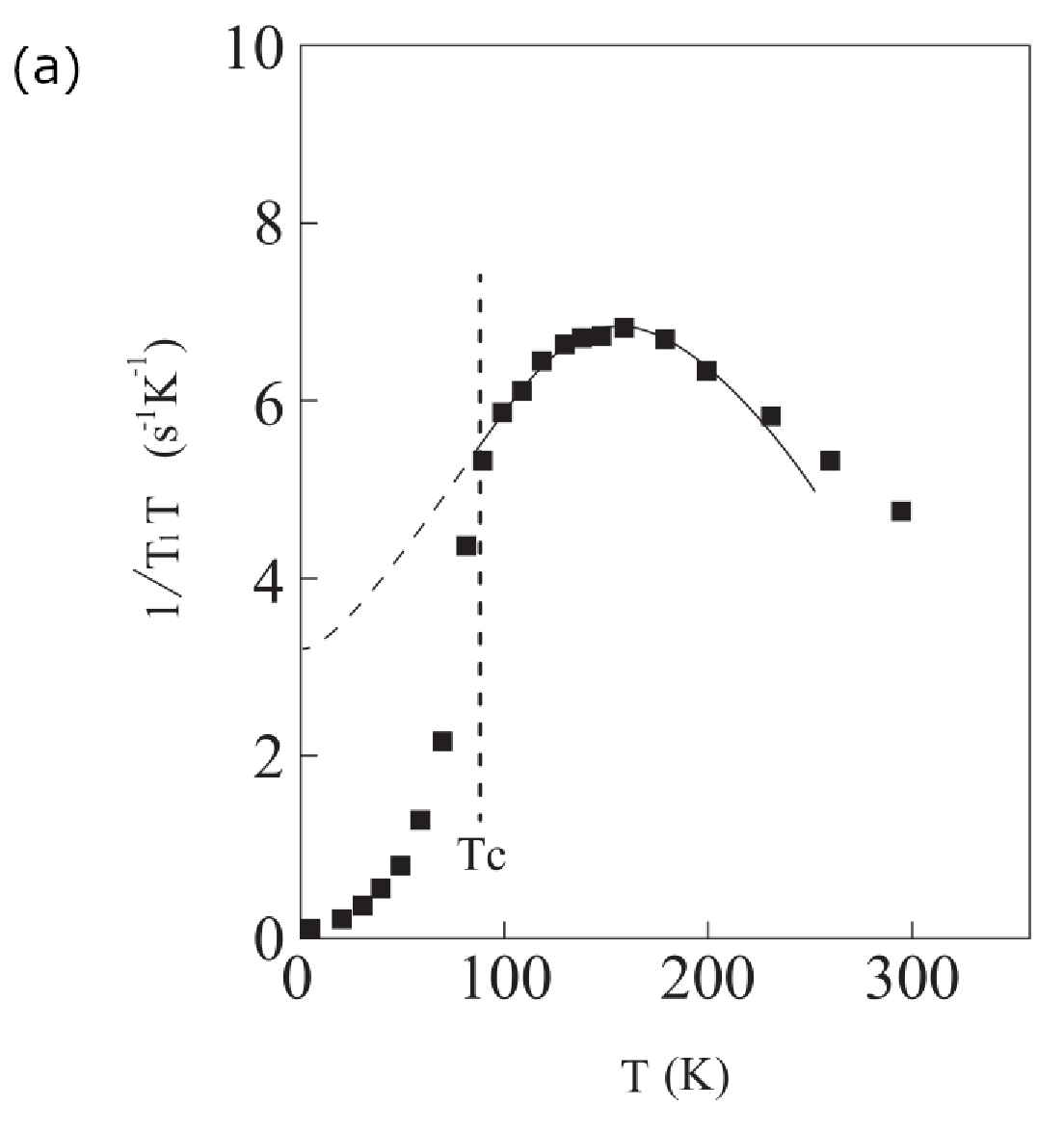}   
\centering 
    \includegraphics[width=\linewidth]{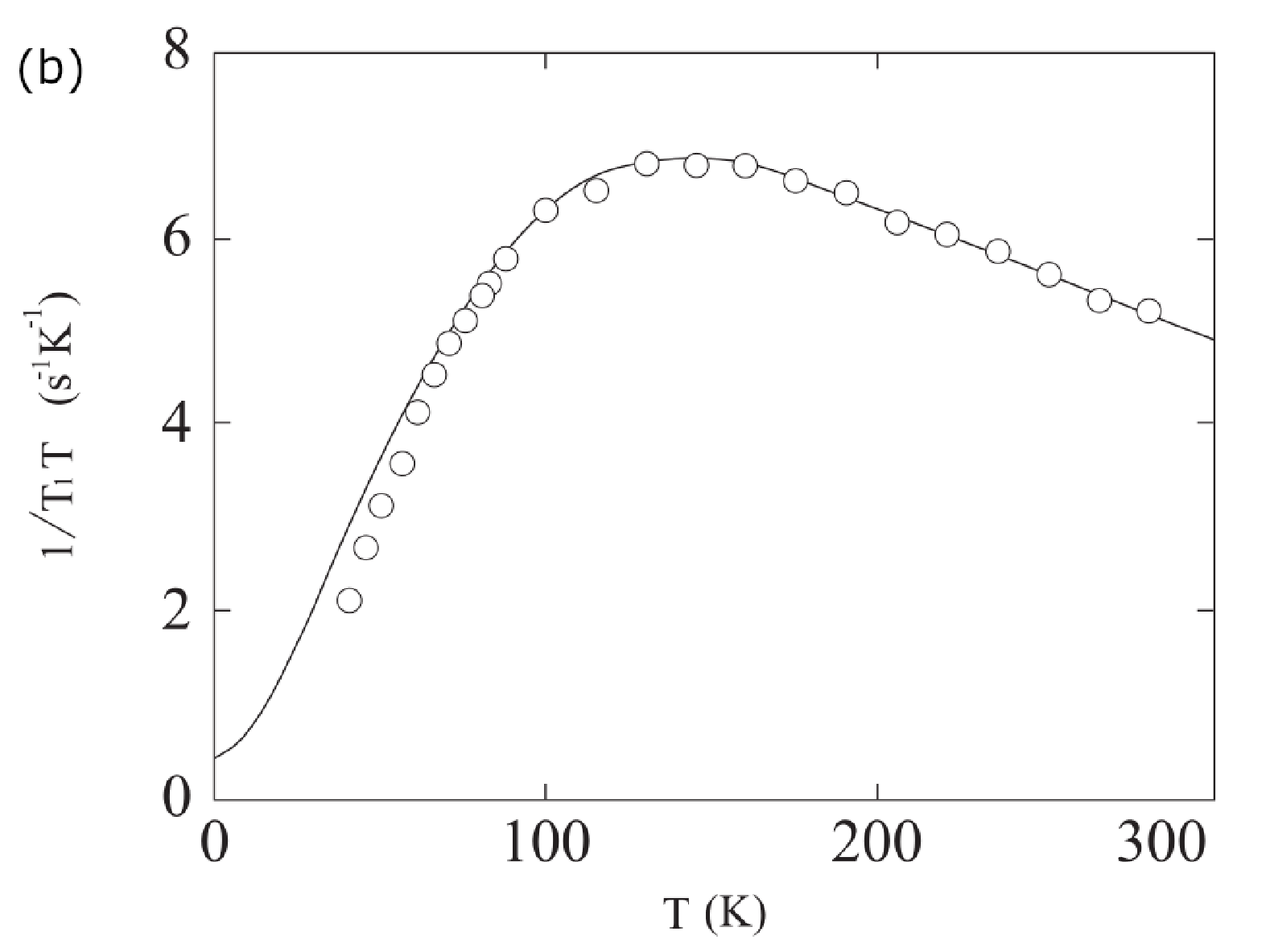}
	\caption{
	Temperature dependence of the nuclear spin-lattice relaxation time of (a) $\mathrm{YBa_2 Cu_4 O_8}$; 
	and (b) $\mathrm{YBa}_2 \mathrm{Cu}_3 \mathrm{O}_{6.64}$; symbols are observed data \cite{12',23}, solid and dashed lines show theoretical fits, Eqs. \eqref{26} and \eqref{26b} in the text.
	}
	\label{fig7'}
\end{figure}

\begin{figure}[!tb]
	\centering
    \includegraphics[width=\linewidth]{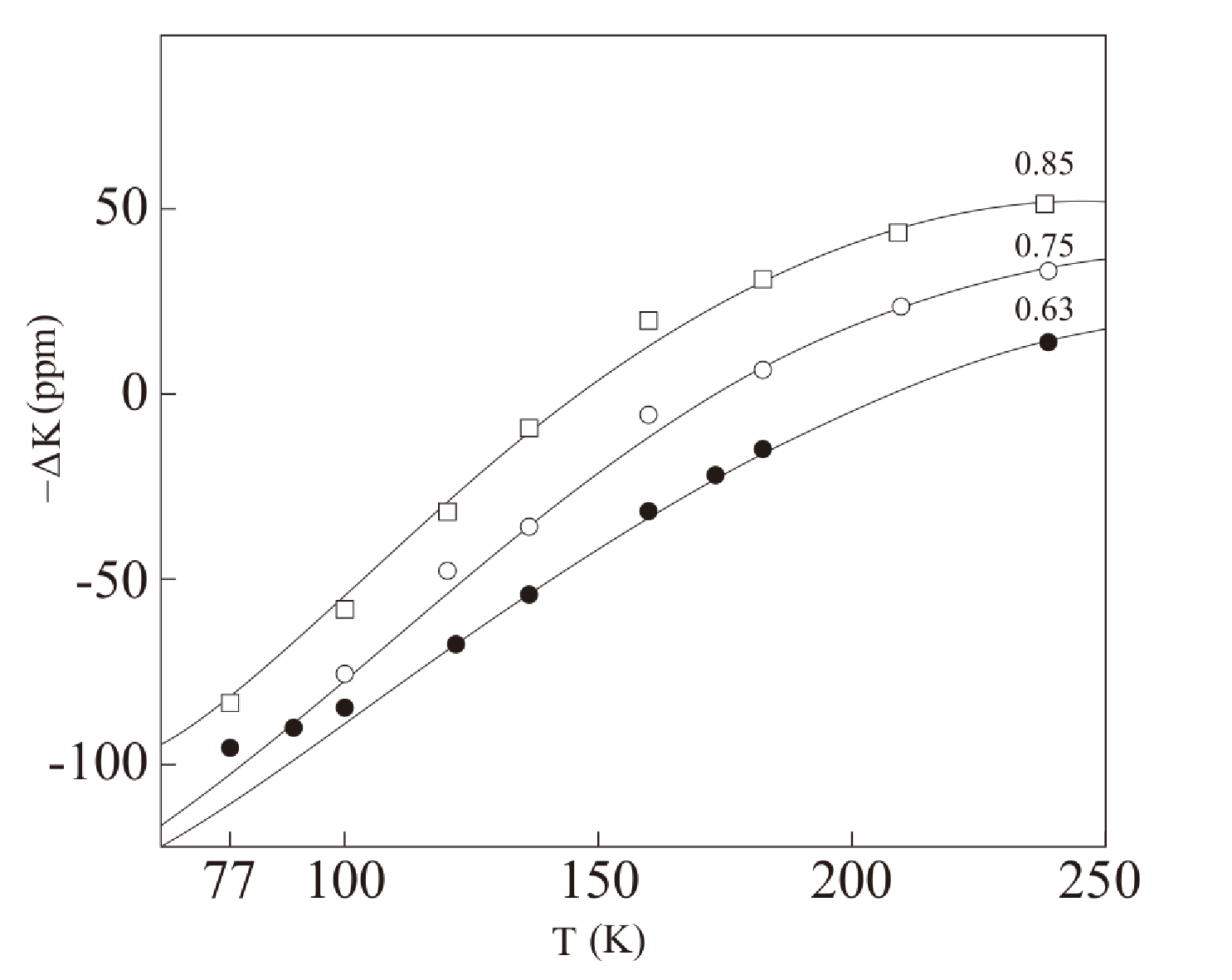}
	\caption{
$\varDelta K(T)$ of three $\mathrm{YBa}_2 \mathrm{Cu}_3 \mathrm{O}_{6+x}$ samples, $x=0.63, \; 0.75$ and $0.85$, for temperatures above $ 77 \; \mathrm{K} $; symbols are experimental data \cite{23}, and solid lines show the theoretical fits; Eq.~\eqref{27'} in the text.
	}
	\label{fig8'}
\end{figure}

It has been a puzzle, for a long time, why LSCO compounds do not exhibit the $ 1/ T_1 T $ maximum.
The key to the puzzle will be found in comparison between two characteristic temperatures;
the temperature of the $1/T_1 T$ maximum, $T_{\rm max}$, and the superconducting transition temperature $T_{\rm C}$. 
The experimental data \cite{12} of $1/ T_1 T$ for another YBCO compound, $\mathrm{YBa_2 Cu_3 O_7}$, above $T_{\rm C}$ ($92 \; \mathrm{K}$) is found to be proportional to $ 1- (T/169)^2 \ln (T/150) $;
the maximum of $1/ T_1 T$ which would appear at $150 / \sqrt{e} = 90.9 \; \mathrm{K}$ is eliminated because of the superconducting transition.
Similarly the non-existence of the $1/ T_1 T$ maximum in LSCO compounds may be explained by this mechanism;
to confirm this experimentally, very precise data above $T_{\rm C}$ are needed.

Alloul et al.~\cite{13'''} have examined the magnetic properties of $\mathrm{YBa}_2 \mathrm{Cu}_3 \mathrm{O}_{6+x} $, $0.35 < x< 1$, through measurements of Knight shift $\varDelta K (T)$ and nuclear spin-lattice relaxation time $T_1$.
We analyse their $\varDelta K(T)$ data and plot in Fig.~\ref{fig8'}; $\varDelta K $ is shown to be expressed in the same form as that predicted from the 3D Fermi liquid model;
\begin{align} \label{27'}
- \varDelta K =  - K_0 \Big\{ 1 + \Big( \frac{T}{T_0} \Big)^2 \ln \frac{T}{T^*} \Big\} \: , 
\end{align}
where $K_0$, $T_0$ and $T^*$ are constants.
For 3 different samples, $x=0.85$, $0.75$ and $0.63$, $K_0$ range as $146, \,162  $, and $156 \; \mathrm{ppm}$, $T_0$ does as $145 , \, 164 $ and $184 \; \mathrm{K}$, and $T^*$ does as $ 394 $, $423$ and $ 457 \; \mathrm{K} $. 
These constants are smoothly varying functions of $x$.
Through analysis of $T_1$, they have found that $(T_1 T)^{-1/2}$ and $ ( - \varDelta K)  $ satisfy the Korringa relation, $(T_1 T)^{-1/2}$ being proportional to $(-\varDelta K)$; this proves that the Fermi liquid description is rightly assured for this system.

In cuprates, the interrelation between the $\mathrm{Cu}(3d)$ holes and $\mathrm{O}(2p)$ holes is of principal concern.
Some authors assist that the $\mathrm{Cu}(3d)$ spins could decouple from the charge transport through $\mathrm{O}(2p)$ holes.
Alloul et al. \cite{23}, however, have found that the $T$-dependence of $\chi(T)$ on oxygen sites
scales the macroscopic susceptibility $\chi_m$ which is dominated by $\chi (T)$ on the copper sites.
The $T$-dependent part of $\chi_m$ should not be attributed solely to the $\mathrm{Cu}^{++}$ spins; 
in conclusion, $\mathrm{Cu}(3d)$ and $\mathrm{O}(2p)$ do not behave as independent degrees of freedom and their hybridization plays a principal role.
This picture coincides with the standpoint of the present paper;
we assume that there exists only one kind of quasiparticles which is specified by $\bm{p}$ vs $\varepsilon(\bm{p})$ relation.

\section{The Logarithmic Temperature Dependence of Transport Properties}

Here we discuss the temperature dependence of transport properties;
i.e. the electrical resistivity, Hall coefficient, and thermoelectric power.

The electrical resistivity arises mainly by the Coulomb scattering and impurity scattering.
In terms of corresponding conductivities $\sigma_{\mathrm{C}}$ and $\sigma_{\mathrm{imp}}$, we have
\begin{align} \label{5.1}
\rho (T) = ( \sigma_{\mathrm{C}} )^{-1}  +  ( \sigma_{\mathrm{imp}} )^{-1} \: .
\end{align}
These conductivities are given, in terms of corresponding relaxation time $\tau (\varepsilon)$, as
\begin{align} \label{5.2}
\sigma =  -  \frac{  2  e^2  }{  3  }  \int \! d \varepsilon \,  v^2  \nu (\varepsilon)  \tau  (\varepsilon)  \frac{ \partial f }{ \partial \varepsilon  }  \: ,
\end{align}
where $e$ is the electronic charge,  $v^2 = ( \partial \varepsilon / \partial p )^2 $, $ f=( e^{\beta \varepsilon} + 1 )^{-1} $.
Here $v^2$ contains, according to Eq.~\eqref{1.2}, a logarithmic term
\begin{align} \label{5.3}
v^2  =  v_0^2
			\bigg\{
			1 +  \mathrm{O} ( \tilde{p} )  -  \frac{ 8 }{ \pi^2 } ( k_{\mathrm{F}}  a )^2  \,  \tilde{p}^2  \ln \frac{  |\,  \tilde{p}  \,|  }{ \tilde{p}_1 }
			 \bigg\} \: ,
\end{align}
where $ v_0 = p_0 / m $, and $ \tilde{p}_1 $ is a constant.
For more general Fermi liquids, since $ \tilde{p} = ( m^{*} / p_0^2 ) \varepsilon  + \mathrm{O} (\varepsilon^2 )  $, $v^2$ behaves as
\begin{align} \label{5.4}
v^2  =  v_0^2  \bigg\{  1  +  \eta_1  \varepsilon  -  \eta_2  \varepsilon^2  \ln  \frac{ |\, \varepsilon \,| }{ \eta^{*} }  \bigg\} \:  ,
\end{align}
where $\eta_1, \eta_2$, and $\eta^*$ are constants.
Concerning $\nu ( \varepsilon )$, we can write, as a generalization of Eq.~\eqref{1.3},
\begin{align} \label{5.5}
\nu (\varepsilon)  =  \nu( 0)  \bigg\{  1  + \nu_1 \varepsilon + \nu_2 \varepsilon^2  \ln  \frac{  | \, \varepsilon \, |  }{ \nu^{*} }  \bigg\} \: ,
\end{align}
where $\nu_1 , \nu_2$ and $\nu^*$ are constants.

Concerning the Coulomb scattering, because of the Umklapp process, $\tau(\varepsilon)$ is given by $\tau_0 (\varepsilon) / T^2$, and the conductivity is given by $ \sigma_{\mathrm{C}} = ( \gamma T^2 )^{-1}  $,  where $\tau_0 ( \varepsilon )$ is  a function of $\varepsilon$ and $\gamma$ is a constant \cite{24}.
If we consider the effect of logarithmic terms in $ v^2 \nu(\varepsilon) $ of Eq.~\eqref{5.2}, there appear a term of the order $T^4 \ln T $ for $\sigma_{\mathrm{C}}$ which will be ignored in this paper.
Thus , we have for the Coulomb resistivity,
\begin{align} \label{5.6}
\rho_{\mathrm{C}}  =  \gamma T^2  \: .
\end{align}

Concerning the impurity scattering, the form of the relaxation time $ \tau (\varepsilon) $ varies with assumptions;
whether the scattering of a quasiparticle with impurities is treated as the single particle scattering, the Born approximation or unitary approximation, $\tau(\varepsilon)$ becomes independent of, inversely proportional to, or proportional to $\nu (\varepsilon)$.
Thus we have
\begin{align} \label{5.7}
v^2 \nu(\varepsilon) \tau(\varepsilon) = v_0^2 \nu(0) \tau(0)
	\bigg\{  1 + \phi_1 \varepsilon + \phi_2 \varepsilon^2 \ln \frac{ |\, \varepsilon\, | }{ \phi^* } \bigg\} \: ,
\end{align}
where $\phi_1$ and $\phi^*$ are constants, and $\phi_2$ takes $ - \eta_2 + \nu_2 $, $- \eta_2$, or $- \eta_1 + 2 \nu_2$ for above three assumptions.

From Eqs. \eqref{5.2} and \eqref{5.7}, $\sigma_{\mathrm{imp}}$ is obtained as
\begin{align} \label{5.8}
\sigma_{\mathrm{imp}} = \frac{ 2 e^2 }{3 } v_0^2 \nu(0) \tau(0)
	\bigg\{
	1 + \frac{ \pi^2 }{ 3 } \phi_2  ( k_{\mathrm{B}} T )^2
			\ln \frac{  T  }{  T_{\mathrm{imp}}  }
	 \bigg\}  \:  ,
\end{align}
where the order $T^2 $ term, $ ( \pi^2 / 3 ) \phi_2' ( k_{\mathrm{B}} T)^2 $, is considered to be included in $ T^2 $ $\ln ( T/ T_{\mathrm{imp}} ) $ term, since
\begin{align*}
\phi_2' (k_{\mathrm{B}} T)^2 + \phi_2 (k_{\mathrm{B}} T)^2 \ln \frac{ T }{ T^* }  =  \phi_2  ( k_{\mathrm{B}} T)^2  \ln \frac{ T }{ T^* e^{ - \phi_2' /\phi_2 } }
\end{align*}
and $T_{\mathrm{imp}} = T^* e^{ - \phi_2' /\phi_2 }$.

When $\sigma_{\mathrm{imp}} (T)$, Eq.~\eqref{5.8}, is abbreviated as $\sigma_0 + \sigma_1 T^2$ $ \ln ( T/ T_{\mathrm{imp}} )$, the total resistivity, Eq.~\eqref{5.1}, is given by
\begin{align} \label{5.9}
\rho (T)
	&= \gamma T^2 + \Big\{ \sigma_0 + \sigma_1 T^2 \ln ( T/ T_{\mathrm{imp}} ) \Big\}^{-1}
\notag \\
	&= \frac{1 }{ \sigma_0 } - \frac{ \sigma_1 }{ \sigma_0^2 } T^2 \ln  \frac{T}{ T_{\mathrm{imp}} } + \gamma T^2  + \mathrm{O} \big( T^4 ( \ln T )^2  \big)
\notag \\
	&= \rho_0 - \rho_1 T^2 \ln \frac{ T }{ T_{\rho}^* } + \mathrm{O} \big( T^4 ( \ln T )^2 \big) \: ,
\end{align}
up to order $T^2$, where $\rho_0 = \sigma_0^{-1}$, $\rho_1 = \sigma_1 / \sigma_0^2$ and $ T_{\rho}^* = T_{\mathrm{imp}} \exp ( \sigma_0^2 \gamma / \sigma_1 ) $.
It should be noted that the $\gamma T^2$ Coulomb resistivity is included in the second term of Eq.~\eqref{5.9}.

Concerning higher order terms with respect to $T$, we should take notice of the difference between the resistivity $\rho (T)$ and the conductivity $\sigma(T)= 1/ \rho (T)$.
Higher order terms of $\rho(T)$, Eq.~\eqref{5.9}, are given by $T^4 \ln T$ and $ T^4 (\ln T)^2 $, while the conductivity has no $ T^4 (\ln T)^2 $ term.
On analysing the experimental data for higher temperatures, the conductivity may be more effective \cite{14}.

If we make the ratio $ \rho_1 / \rho_0 $, since $ \rho_1 / \rho_0 = \sigma_0 / \sigma_1 = ( \pi^2 / 3 ) k_{\mathrm{B}}^2 \phi_2 $, it does not depend on the character of impurities, but reflects solely the nature of the electron system; concerning this, measurements done by Ni et al \cite{25} for the iron-based superconductor, $\mathrm{Ba(Fe_{1-{\it x}} Co_{{\it x}} )_2 As_2 }$, is instructive.
They cut a single crystal of $ x = 0.038 $ sample into three parts and measured three different $\rho (T)$'s.
We have analysed their data to find $\rho_1, \rho_2$ and $\rho_3$ for 3 parts in units of $\mathrm{m\Omega\cdot cm}$;
\begin{align} \label{5.10}
\rho_1 (T) &= 0.887 \{ 1 - 3.67 \times 10^{-6} T^2 \ln ( T / 2180 ) \} \: ,
\notag \\
\rho_2 (T) &= 0.503 \{ 1 - 4.90 \times 10^{-6} T^2 \ln ( T / 1220 ) \} \: ,
\\
\rho_3 (T) &= 0.234 \{ 1 - 3.68 \times 10^{-6} T^2 \ln ( T / 2420 ) \} \: .
\notag
\end{align}
Here, although $\rho (0)$ values are all different, the relative $T$-variations of $\rho_1$ and $\rho_3$ are almost the same.
In particular, the fact that $\rho_1/ \rho_0$ values, $3.67$ and $3.68  \times 10^{-6}$, are almost exactly the same shows the proof for our statement that $\rho_1/ \rho_0$ values do not depend on impurities.

In spite of the preceding statement, the $T$-dependence of $\rho_1$ and $\rho_2$ is quite different.
Even in the single crystal, the distribution or quantity of impurities and lattice defects varies with positions. 
Thus we should note that the molecular theoretical calculation for transport properties based on the first principle may not be very effective.

Next we shall describe the $T$-dependence of the Hall coefficient $R_{\mathrm{H}} (T)$.
If the system is composed of one kind of carriers (quasiparticles), $R_{\mathrm{H}} (T)$ is defined as
\begin{align} \label{5.11}
R_{\mathrm{H}} ( T) = \frac{ e \tau(\mu)}{ mc} \rho (T) \: ,
\end{align}
and its temperature dependence follows that of $\rho(T)$.
If the system is two band metals with equal number of electrons and holes, $R_{\mathrm{H}}$ is proportional to $ ( \rho_e - \rho_h ) / ( \rho_e + \rho_h ) $, where $ \rho_e$ and $\rho_h$ are resistivity functions for electrons and holes.
Since each resistivity can be expanded in powers of $T^2$ and $ T^2 \ln T $, the resultant $R_{\mathrm{H}}$ can be expressed in the $ a-b T^2 \ln ( T/ T^*_{\mathrm{H}} ) $ form, $a$, $b$ and $T_{\mathrm{H}}^*$ being constant.

Finally, the temperature dependence of the thermoelectric power, $Q (T)$, will be described.
When A. H. Wilson wrote the book ``Theory of Metals'', in 1953, he listed the data of $Q(T)$ for 3 metals including $\mathrm{Pt}$.
Facing the data, he wrote ``The behaviour at very low temperature is still more complicated and cannot at the moment be reconciled with the theory''.
The inverse-$v$ shape structure of $Q$ vs $T$ curve at low temperatures was not able to be explained at that time.

Since $Q(T)$ is proportional to an average of $\varepsilon$ given by
\begin{align} \label{5.12}
Q =  -  ( e T )^{-1}
	\frac{
	\int \! d \varepsilon \, \varepsilon v^2 \nu(\varepsilon) \tau(\varepsilon) \partial f / \partial \varepsilon
	  }{
	  \int \! d \varepsilon \,  v^2 \nu(\varepsilon) \tau(\varepsilon) \partial f / \partial \varepsilon
	  }
	  \: ,
\end{align}
here we need to consider higher order term in Eq.~\eqref{5.7},
\begin{align} \label{5.13}
v^2 \nu( \varepsilon ) \tau ( \varepsilon ) = v_0^2 \nu_0 \tau_0
	\bigg\{
	1 + \phi_1 \varepsilon + \phi_2 \varepsilon^2 \ln \frac{ |\, \varepsilon \, | }{ \phi^* }  +  \phi_3  \varepsilon^3 \ln \frac{ |\, \varepsilon \, | }{ \phi^{**} }
	\bigg\} \: ,
\end{align}
where $\phi_3$ and $\phi^{**}$ are constants.
After taking the average, $Q$ is obtained as
\begin{align} \label{5.14}
Q(T) = - \frac{ \pi^2 }{ 3 } \frac{ k_{\mathrm{B}}^2 }{ e } \phi_1 T
	\bigg\{
	1 + \bigg( \frac{ 7 \phi_3 }{ 5 \phi_1 }  - \frac{3}{2} \phi_2 \bigg) \pi^2 ( k_{\mathrm{B}} T )^2 \ln \frac{ T }{ T_{Q}^* }
	\bigg\} \: ,
\end{align}
up to order $T^3$, where $T_{Q}^* $ is a constant which depends on $ \phi^* $, $ \phi^{**} $, $\phi_2$ and $\phi_3$.
For higher temperatures $ T^5 \ln ( T/ T^* ) $ term is needed.

Thus $Q(T)$ takes the form  $ q_1 T + q_2 T^3 \ln ( T/ T_{Q}^* ) $, where $q_1, q_2 $ and $T_Q^*$ are constant; as a function of $T$, $Q(T)$ behaves as $v$-shape, or inverse $v$-shape, and has a minimum, or maximum.
We have already shown that the maximum appears in prototype heavy fermion compound $\mathrm{CeAl_3}$ \cite{13} and the minimum does in the normal state of iron-based superconductors \cite{2}.

\section{Electrical resistivity of LSCO and YBCO}

\begin{figure}[!bht]
	\centering
    \includegraphics[width=\linewidth]{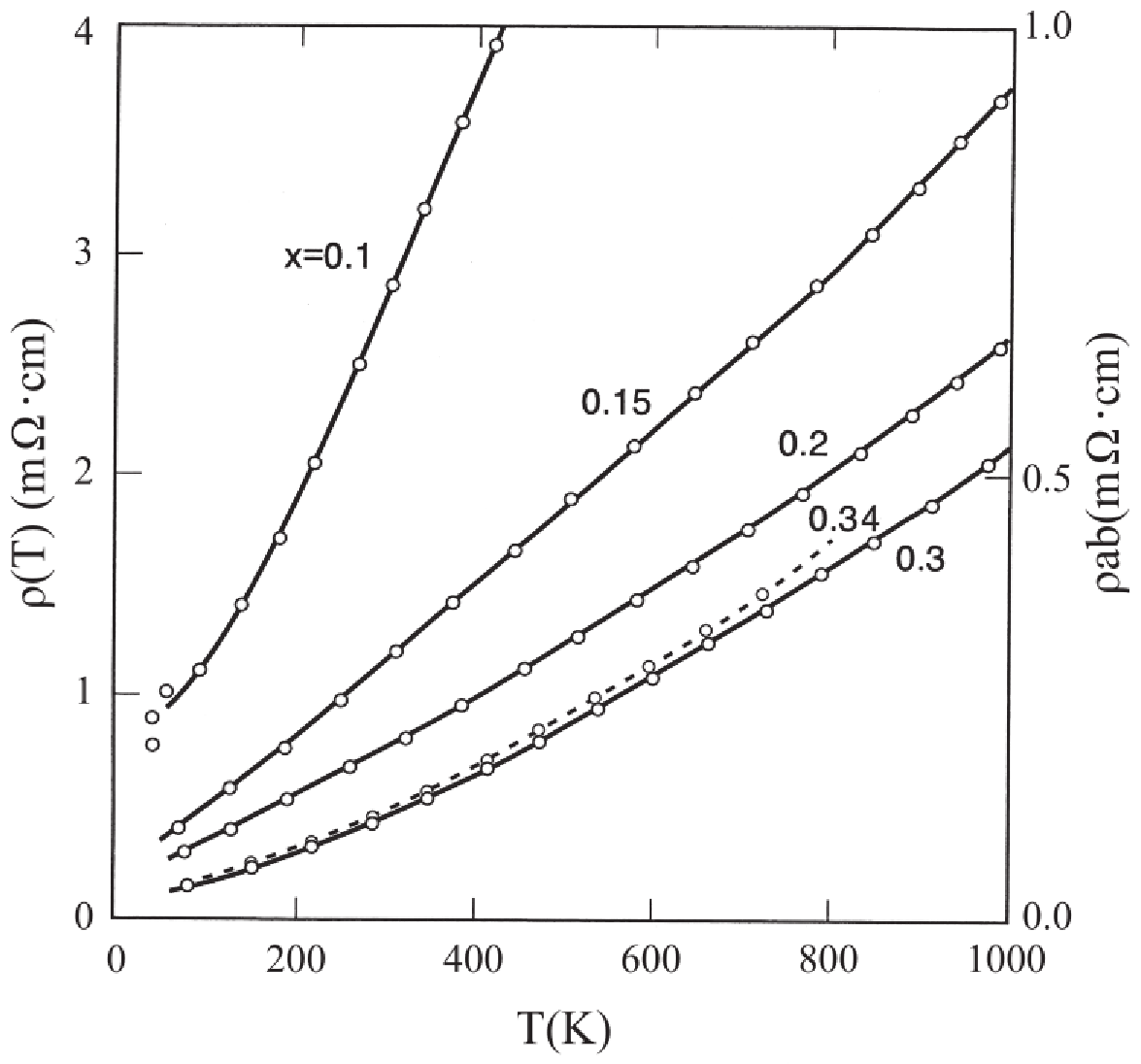}
	\caption{
$\rho (T)$ vs $T$ of single crystal ($x=0.34$ right scale) and polycrystalline $ \mathrm{La}_{2-x} \mathrm{Sr}_{x} \mathrm{Cu}\mathrm{O}_{4} $ samples (left scale); circles are representative points of experimental data \cite{26}. 
A dashed line and solid lines are theoretical fit for single crystal and polycrystalline samples. 
Labels show $x$-values.
	}
	\label{fig7}
\end{figure}

Takagi and his group \cite{26} have performed extensive measurements of the in-plane electrical resistivity, $\rho_{ab}$, of high quality LSCO single crystals and polycrystalline samples for the composition range, $ 0 < x \leq 0.34 $, and for temperatures up to $ 1000 \, \mathrm{K} $;a part of their results is shown in Fig.~\ref{fig7}.
They also have made $ d \rho / d T $ vs $T$ measurements on the same samples.

Their main interest is to examine the validity of the power law formula, $ \rho = \rho_0 + A T^n $, and to clarify the nature of the $T$-linear resistivity on the ideas of 2D strongly correlated electron systems.
They have concluded $ n= 1.5 $ for $ x = 0.34 $ single crystal sample.
We reexamine their result;
we choose 7 representative points in the $ \rho (T) $ curve for the range $ T \leq  700 \mathrm{K} $, carefully read the points and try to find the best fitting value of $n$.
We have selected 5 $n$-points from $n=1.3$ to $1.5$ and evaluate the $rms$ error ($\delta$) of the fitting in $\%$, as a function of $n$, to find $ \delta = 0.475 + 81.2 ( n- 1.400 )^2 $;
the best fitting value of $n$ is $1.40$, the error being $0.48 \%$, and the fitting formula is given by
\begin{align} \label{6.1}
\rho(T)  = 0.0123 \, \bigg\{ \,  1  +  \bigg(  \frac{  T  }{  63.7  }  \bigg)^{1.40} \;  \bigg\} \; \mathrm{m\Omega \cdot cm } \: .
\end{align}
Correspondingly, the resistivity formula for the $T^2 \ln T$ variation is found to be
\begin{align} \label{6.2}
\rho(T)  = 0.0212 \, \bigg\{ \, 1  -  \bigg(  \frac{  T  }{  216  }  \bigg)^{2} \ln \frac{ T }{ 3280 } \;  \bigg\} \; \mathrm{m\Omega \! \cdot \! cm } \: ,
\end{align}
and the error is $0.16 \%$ which is just $1/3$ of the $T^{1.40}$ law.
The $T^2 \ln T$ law reproduce the experimental data the most accurately.

We already analysed the main part of their data and published the result in 2003 \cite{3}; four $\rho (T)$ curves for $ x = 0.1, \, 0.15 , \, 0.2 $ and $ 0. 3 $ polycrystalline samples were analysed.
Here we restate the essential points.

For the analysis of $ \rho(T) $ we should  note that the observed data include the resistivity due to the electron-lattice interactions;
this resistivity starts from the $ T^5 $ law at lowest temperatures,
the power $5$ decreasing with increasing temperature, and finally reaching to $ \varDelta \rho \propto T $ for $T \gtrsim 0.2 \, \Theta_{\mathrm{D}}$ , where $ \Theta_{\mathrm{D}} $ is the Debye temperature.
We have already mentioned that the $T$ linear part of $ \rho (T) $ can be approximately expressed by $ T^2 \ln ( T/ T^{*} ) $ function;
hence, the fitting data on the basis of $ \varDelta \rho \sim T^2 \ln T $ law may vary with the range of temperature.

Fitting procedure for $ x= 0.15 $ sample is as follows;
for temperature up to $ 400 \, \mathrm{K} $, the resistivity is shown to follow
\begin{align} \label{6.3}
\rho (T)  = 0.273 \bigg\{  1 - \bigg(  \frac{ T }{ 200.5 } \bigg)^2  \ln \frac {T}{ 1216 }   \bigg\}  \; \mathrm{m \Omega \cdot cm } \: .
\end{align}
When the temperature range is extended to $ 1000 \, \mathrm{K} $, we have to consider $T^4$, $ T^4 \ln T $ and $ T^4 ( \ln T )^2 $ terms;
without $ T^4 ( \ln T )^2 $ term, the resistivity follows
\begin{align}
\rho(T) =  0.249  \bigg\{ & 1 -  \bigg(  \frac{ T }{ 163}  \bigg)^2  \ln \frac{T}{ 748 }  
\notag 
\\
& \:\: -  \bigg(  \frac{ T }{ 456 }  \bigg)^4  \ln \frac{T}{ 2945 }  \bigg\} \; \mathrm{ m \Omega \cdot cm } \: .
 \label{6.4}
\end{align}
with $ \delta = 0.20 \%$, while, with that, it does
\begin{align} \label{6.5}
\rho (T) = 0.2655 \bigg\{  1  &  - \bigg(  \frac{ T }{ 200 }  \bigg)^2  \ln \frac{ T }{ 1490 }  +  \bigg(  \frac{ T }{ 227 }  \bigg)^4 \ln \frac{ T }{ 33.4 }
\notag \\
& - \bigg( \frac{ T }{ 440 } \bigg)^4 ( \ln T )^2    \bigg\} \; \mathrm{ m \Omega \cdot cm } \:
\end{align}
with $\delta = 0.19 \%$.
If we compare these two formulae, the accuracy is the same;
however, the principal form of the low temperature formula is to be continued up to higher temperatures, the inclusion of the $ T^4  ( \ln T )^2 $ term seems to be essential.
Thus Eq.~\eqref{6.5} is our final result.
The same argument has applied to $ x= 0.2 $ and $ 0.3 $ samples.

\begin{figure}[h]
	\centering
    \includegraphics[width=\linewidth]{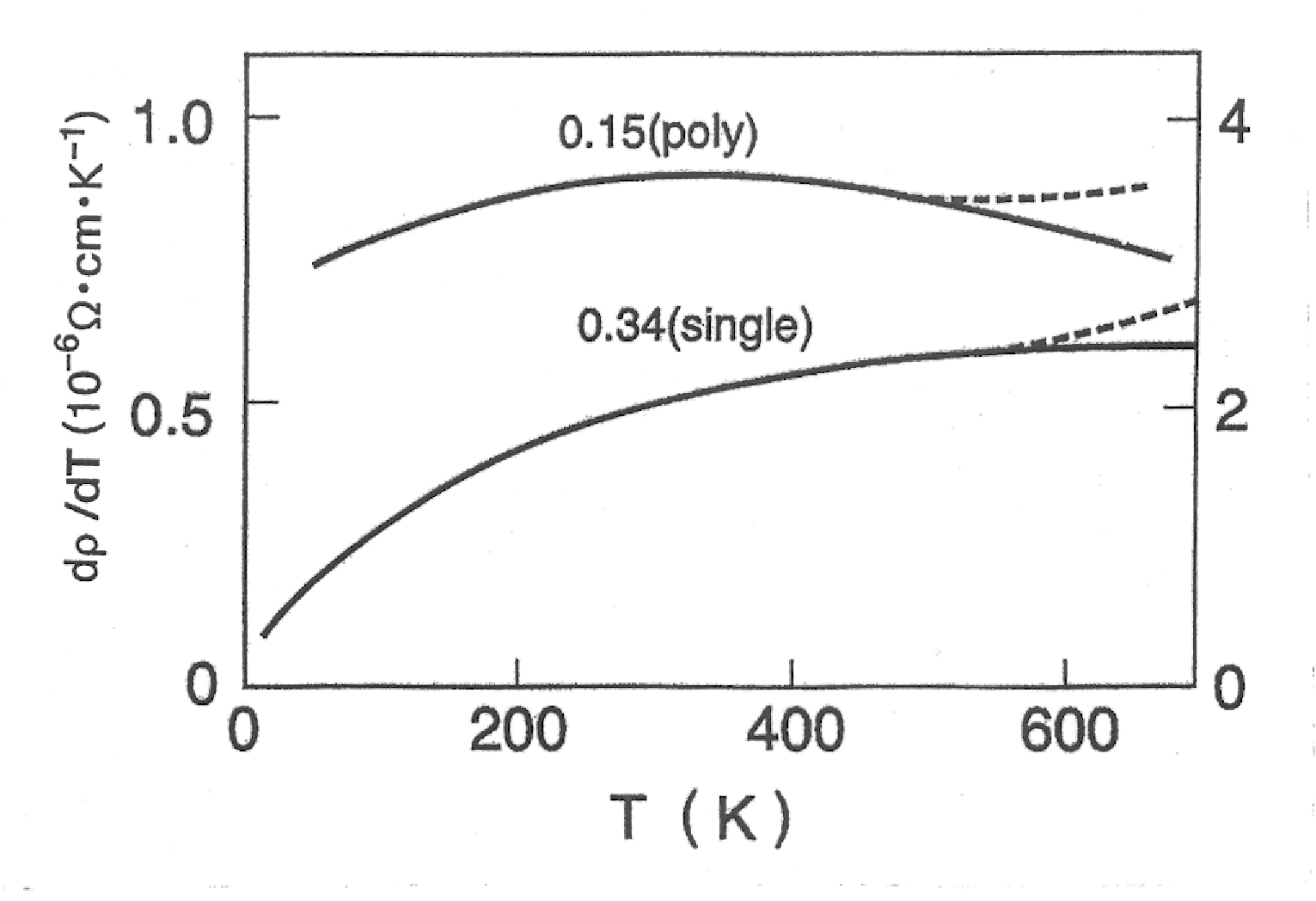}
	\caption{
	$d \rho / d T$ vs $T$ of $ x= 0.15 $ polycrystalline (right scale) and $0.34$ single crystal (left scale) $ \mathrm{La}_{2-x} \mathrm{Sr}_x \mathrm{Cu} \mathrm{O}_4 $ samples.
	The smoothed experimental data \cite{26} fit theoretical $T \ln T$ curves, Eqs.~\eqref{6.6} and \eqref{6.8}, up to about $500 \,\mathrm{K}$ above which the data (dashed lines) deviate from the solid lines.
	}
	\label{fig8}
\end{figure}

The $ T^2 \ln T $ variation of $ \rho (T) $ can be confirmed by analysing the $ d \rho / d T $ vs $T$ diagrams shown in Fig. \ref{fig8}.
It is seen that the observed data can be precisely fitted to the $T \ln T $ variation up to about $ 500 \; \mathrm{K}$; for $ x= 0.15$ polycrystalline sample, the curve can be fitted, excluding the constant term, by
\begin{align} \label{6.6}
\frac{ d \rho }{ d T } = - 1.445 \times 10^{-5} T \ln \frac{ T }{ 876 } \: \: \mathrm{ \Omega \cdot cm \cdot K^{-1} }
\: .
\end{align}
This is to be compared with $ d \rho / d T $ derived from Eq.~\eqref{6.3};
\begin{align} \label{6.7}
\frac{ d \rho }{ d T } = - 1.36 \times 10^{-5} T \ln \frac{ T }{ 737 } \:\: \mathrm{ \Omega \cdot cm \cdot K^{-1} } \: .
\end{align}
For $ x= 0.34 $ single crystal, we find
\begin{align} \label{6.8}
\frac{ d \rho }{ d T } = - 8.77 \times 10^{-7} T \ln \frac{ T }{ 1820 } \:\:
\mathrm{ \Omega \cdot cm \cdot K^{-1} } \: ,
\end{align}
while, from $ \rho (T) $, Eq.~\eqref{6.2}, we obtain
\begin{align} \label{6.9}
\frac{ d \rho }{ d T } = - 9.09 \times 10^{-7} T \ln \frac{ T }{ 1990 } \:\:
\mathrm{ \Omega \cdot cm \cdot K^{-1} } \: .
\end{align}
The differences of two expressions, Eq.~\eqref{6.6} and \eqref{6.7}, and,  Eq.~\eqref{6.8} and \eqref{6.9}, may be caused by difficulties of $ d \rho / d T $ measurements.

We analyse next the experimental data for the electrical resistivity of well annealed samples of $\mathrm{ Y Ba_2  Cu_3 O_{7-{\it x}} } $ by Nakazawa and Ishikawa \cite{27}.
Because of high transition temperatures for superconductivity, to clarify the nature of this compounds seems to be important.


\begin{figure}[!th]
	\centering
    \includegraphics[width=\linewidth]{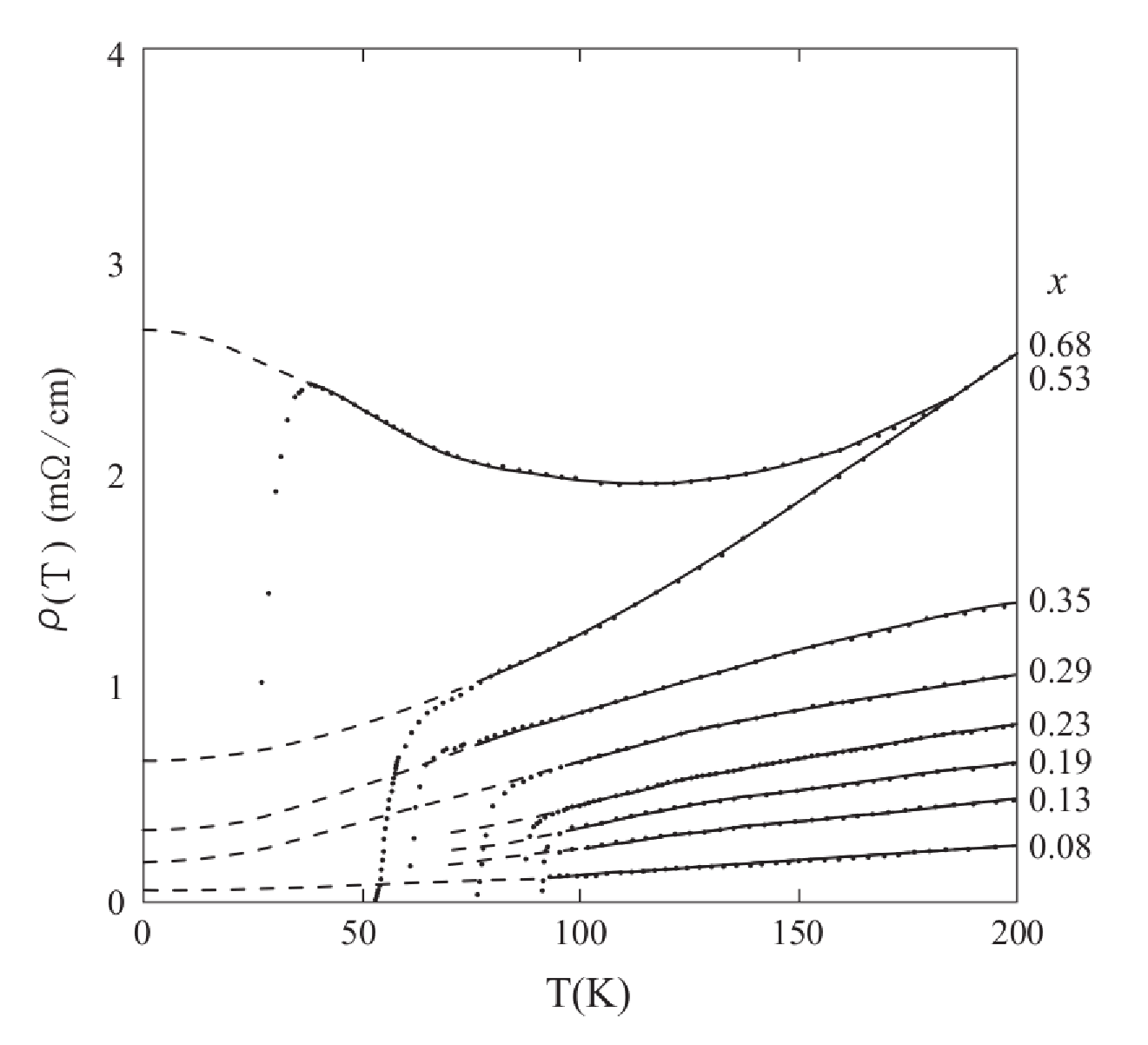}
	\caption{
	$\rho(T)$ vs $T$ of $\mathrm{ Y Ba_2  Cu_3 O_{7-{\it x}} } $ samples.
	Dots show the experimental data \cite{27}; solid lines show the theoretical fit, Eq.~\eqref{6.10} etc. in the text; for reference, the extensions of theoretical curves for lower temperatures are shown by dashed lines.
	}
	\label{fig9}
\end{figure}


Their resistivity data are reproduced in Fig.~\ref{fig9}.
For samples $ x \gtrsim 0.78 $, not shown in the figure, the resistivity increases indefinitely with decreasing temperature because of the variable range hopping type conduction.
For $ x= 0.68 $ sample, however, the increase of $ \rho (T) $ with decreasing $T$ is not due to this hopping mechanism, but is caused by the negative sign of $ \rho_1 $ in the formula, $ \rho_0 - \rho_1 T^2 \ln ( T/ T_{\rho}^* ) $; in fact, the resistivity for $T \lesssim 200 \, \mathrm{K}$ is
\begin{align} \label{6.10}
\rho(T) = 2.59 \bigg\{  1 + \bigg(  \frac{ T }{ 177 }  \bigg)^2  \ln \frac{T}{ 206 }  \bigg\} \: \; \mathrm{ m \Omega \cdot cm } \: ,
\end{align}
while, the resistivity for $ x = 0.53 $ sample is
\begin{align}
\rho(T) =  0.654  \bigg\{  1 - \bigg(  \frac{ T }{ 219 }  \bigg)^2  \ln \frac{T}{7650}  \bigg\} \: \; \mathrm{ m \Omega \cdot cm } \: ;
\end{align}
the sign change occurs at $ x= 0.57 $.
Finally, for $ x = 0.08 $ sample, $\rho(T)$ becomes
\begin{align} \label{6.12}
\rho(T) = 0.0603 \bigg\{  1 - \bigg(  \frac{ T }{ 125 }  \bigg)^2 \ln \frac{ T}{710}  \bigg\} \; \:  \mathrm{ m \Omega \cdot cm }  \:  .
\end{align}
%


\begin{figure}[h]
	\centering
    \includegraphics[width=\linewidth]{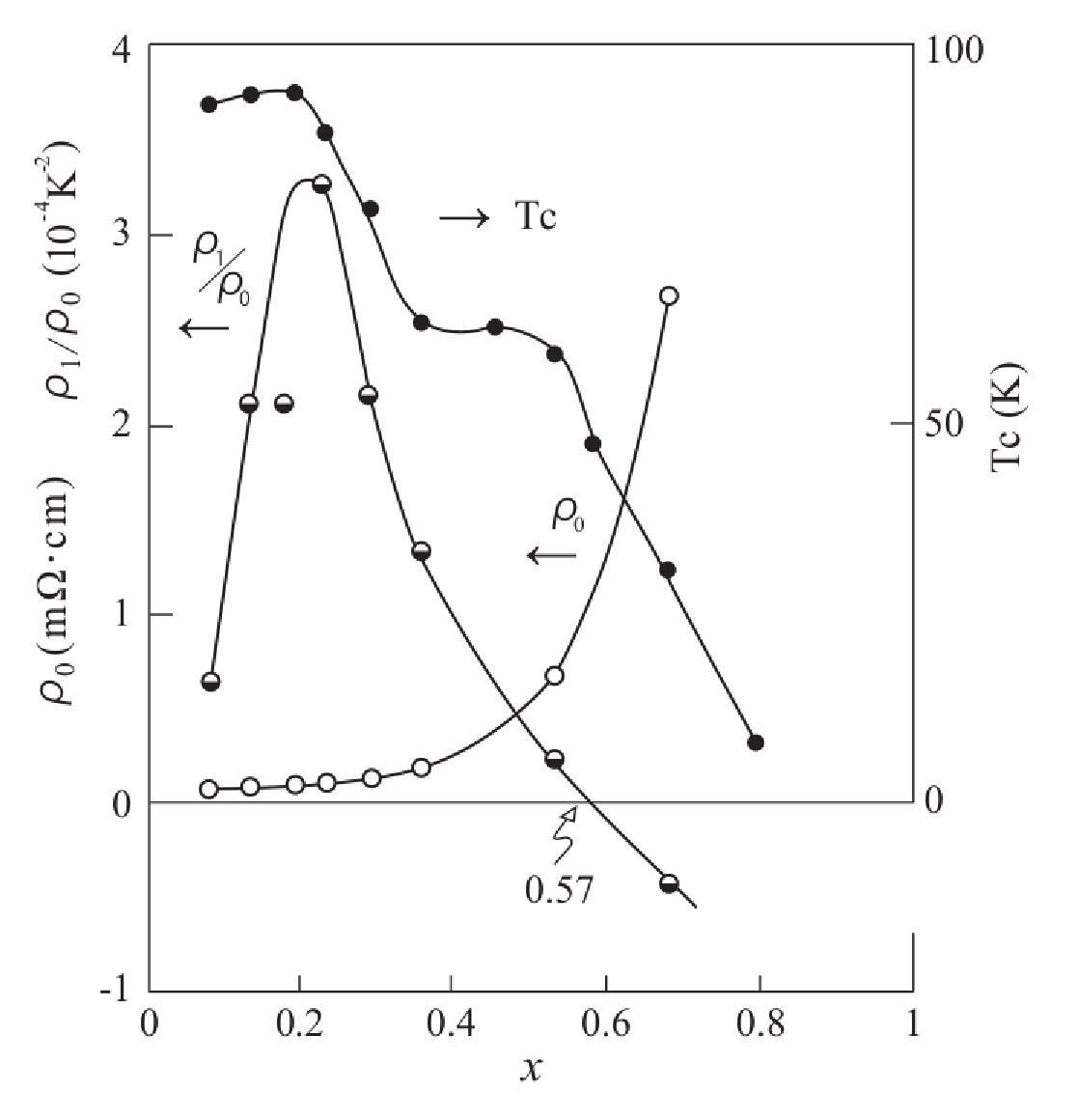}
	\caption{
	$\rho_1 / \rho_0 $ and $T_{\mathrm{C}}$ vs $x$ diagram of $\mathrm{YBa}_2 \mathrm{Cu}_3 \mathrm{O}_{7-x}$ samples; for reference $\rho_0 $ is added.
	$\rho_1 / \rho_0 $ changes the sign at $x= 0.57$.
	}
	\label{fig10}
	\centering
    \includegraphics[width=\linewidth]{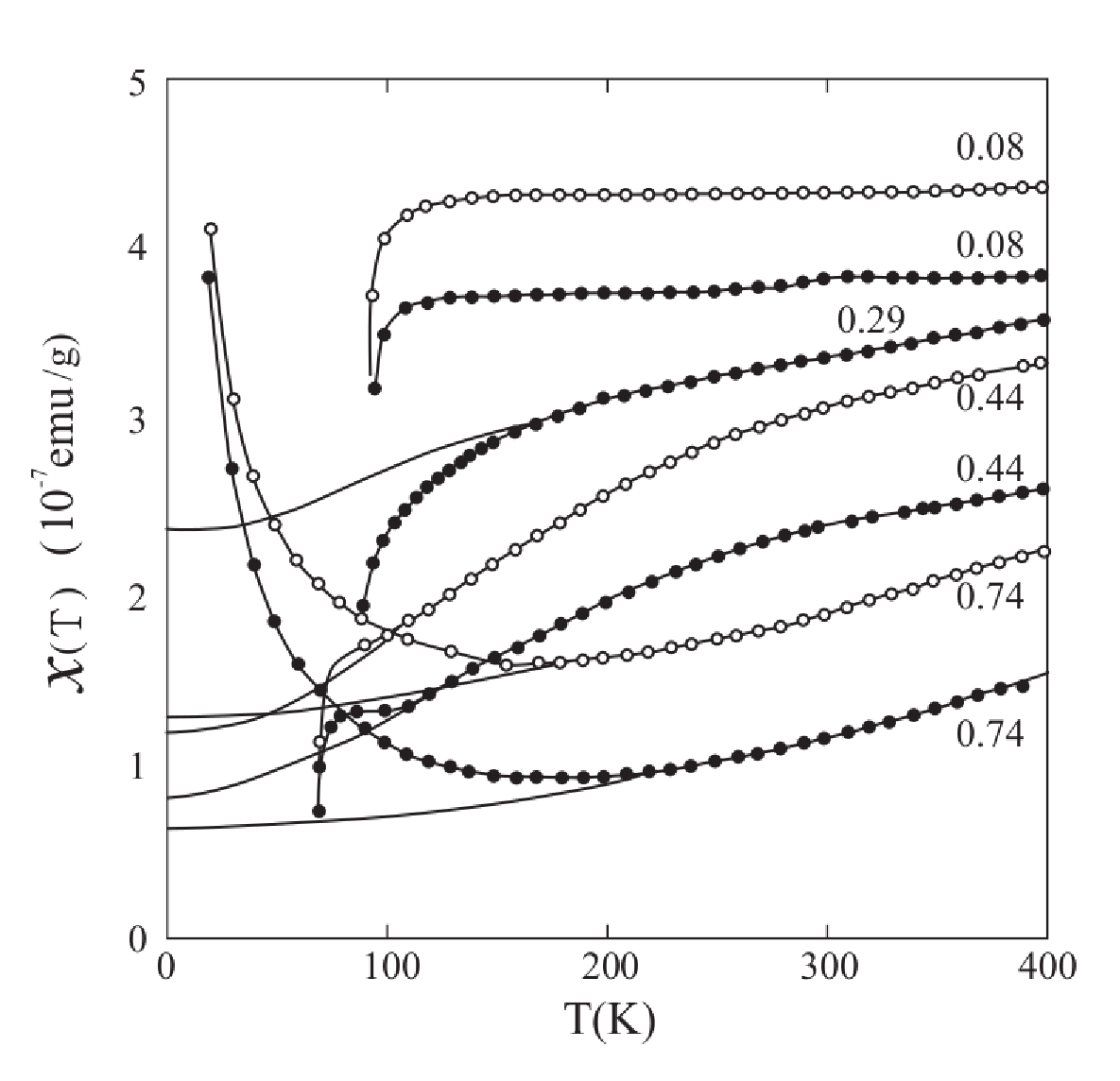}
	\caption{
	 $\chi (T) $ vs $T$ curves of $\mathrm{YBa}_{2} \mathrm{Cu}_3 \mathrm{O}_{7-x}$ samples.
	 Open or filled circles show the experimental data when the applied field is parallel or perpendicular to the $ab$-plane \cite{27}.
	 Solid lines at higher temperature are theoretical fits which are extended up to $0 \, \mathrm{K}$ for reference.
	}
	\label{fig11}
\end{figure}



To examine closely the $x$-dependence of the constants, we plot in Fig.~\ref{fig10}, $T_{\mathrm{C}}$, $\rho_{0} $, and $ \rho_1 / \rho_0 $ as functions of $x$.
Here we can draw the following three points; (i) for $ 0 \leq x \lesssim 0.2 $, $T_{\mathrm{C}} $ is not influenced by $ \rho_1 / \rho_0 $ values, (ii) when $x$ exceeds $0.2$, both $ T_{\mathrm{C}} $ and $ \rho_1 / \rho_0 $ begin to decrease with $x$, (iii) when $x$ exceeds the point of changing the sign of $ \rho_1 / \rho_0 $, $ x \approx 0.5 $, $T_{\mathrm{C}}$ drops rapidly, and $\rho_0$ increases steeply with $x$.
Physical reasons for these features are not clear at present.

Nakazawa and Ishikawa have also measured the magnetic susceptibility of $ \mathrm{ Y Ba_2  Cu_3 O_{7-{\it x}} }  $.
In relation to the electrical resistivity, the $ T $-dependence of $ \chi(T) $ will be discussed here.
Since the $T$-dependence can be classified mainly into three regions of $x$, we reproduce, as representative, the susceptibility data of $ x= 0.08,$ $ 0.29 ,\; 0.44 $, and $ 0.74 $ samples in Fig. \ref{fig11}.
First, concerning $ x = 0.08 $ sample, the temperature variations of both $ \chi(T) $, i.e. $ b $ in Eq.~\eqref{4.4}, and $ \rho(T) $, i.e. $ \rho_1 $ in  Eq.~\eqref{6.12}, are almost zero; this is satisfied only when the interaction strength $ (k_0 a)^2 $ is nearly zero.
Thus the quasiparticles of this system are very weakly interacting;
here it is not known why this weakly interacting system has the highest $T_{\mathrm{C}}$ value.
For intermediate region, $ 0.30 \lesssim x \lesssim 0.50 $, the variation of $ \chi(T)  $ is rather large; here $ \chi(T) $ can be precisely fitted by the $ T^2 \ln T $ variation.
It is shown that, for $ x = 0.44 $ sample, when an applied field is parallel or perpendicular to the $ab$-plane, $ \chi (T) $ is given by
\begin{align}
\chi (T)  = 0.828 \bigg\{ 1 -  \bigg(  \frac{ T }{ 175 }  \bigg)^2  \ln \frac{ T }{ 568 }  \bigg\}  \times 10^{-7}  \: \mathrm{emu/ gr} \: ,
\end{align}
or
\begin{align}
\chi (T)  = 1.180 \bigg\{ 1 -  \bigg(  \frac{ T }{ 188 }  \bigg)^2  \ln \frac{ T }{ 561 }  \bigg\}  \times 10^{-7}  \: \mathrm{emu/ gr} \: .
\end{align}
Here, to be remarked, the relative temperature variations are almost the same.
Finally, for the range $ x \gtrsim 0.70 $, the susceptibility increases indefinitely with decreasing temperature for $ T \lesssim 150 \; \mathrm{K} $, and the authors have stated that $ \chi(T) $ follows the Curie-Weiss law.
Since the system obeys the variable range hopping conduction, it does not behave as  a free Fermi liquid; when the temperature is raised, however, it should become  an itinerant Fermi liquid.
Actually, for $ x = 0.74 $ sample and $ T \gtrsim 200 \;\mathrm{K} $, $ \chi (T) $ is shown, for a field parallel or perpendicular to the $ab$-plane, to follow the $T^2 \ln T $ variation
\begin{align}
\chi(T) = 0.627 \bigg\{  1 - \bigg( \frac{ T }{ 749 }  \bigg)^2 \ln \frac{ T }{ 57160 } \bigg\} \times 10^{-7} \; \mathrm{emu/gr} \: .
\end{align}
or
\begin{align}
\chi(T) = 1.28 \bigg\{  1 - \bigg( \frac{ T }{ 764.5 }  \bigg)^2 \ln \frac{ T }{ 6040 } \bigg\} \times 10^{-7} \; \mathrm{emu/gr} \: ,
\end{align}
In conclusion, the quasiparticles of YBCO form strictly a 3D Fermi liquid.

\begin{figure}[!bh]
	\centering
    \includegraphics[width=0.4\textwidth]{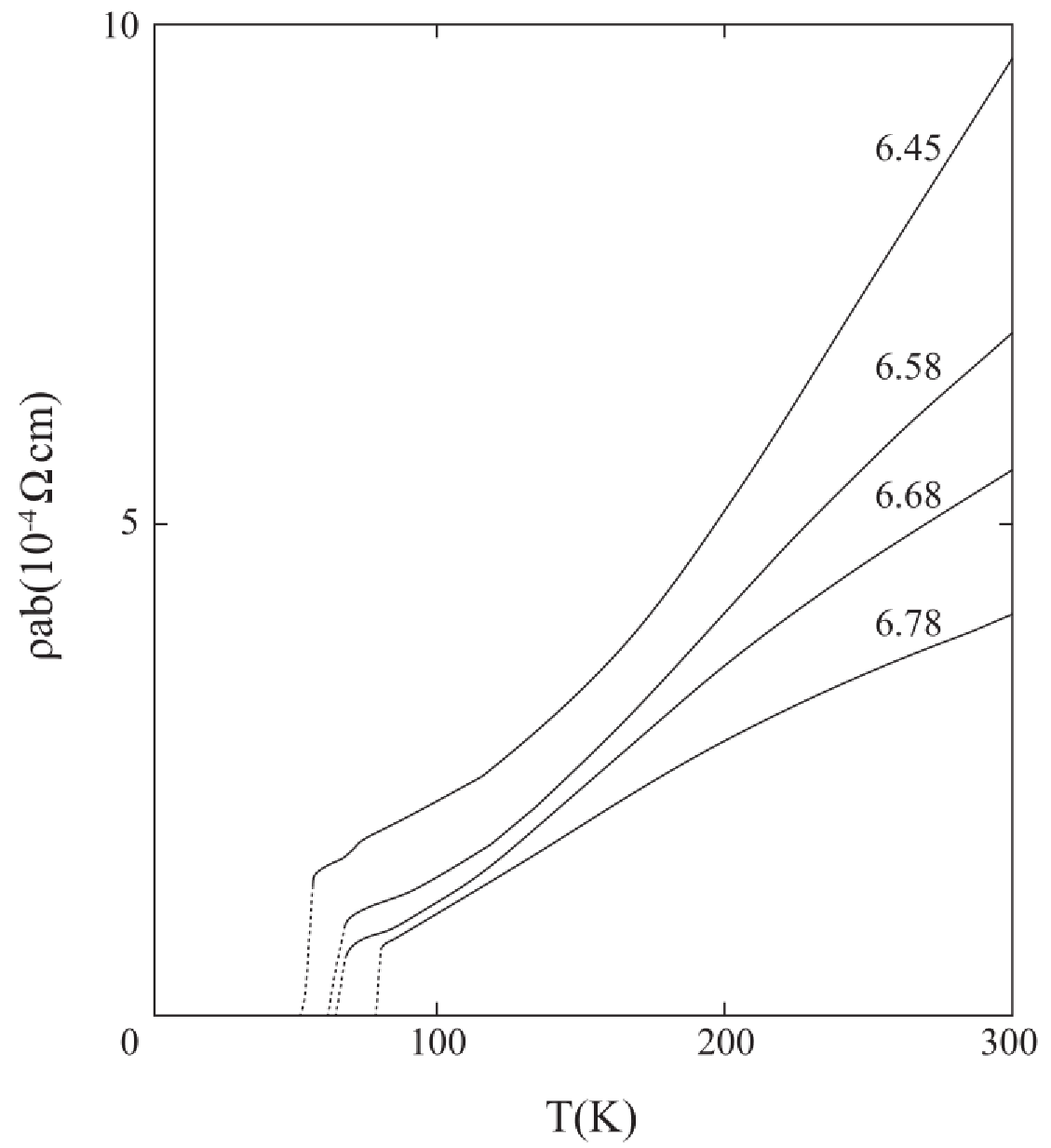}
	\caption{
	The electrical resistivity $ \rho_{ab}(T) $ of $ \mathrm{YBa}_{2} \mathrm{Cu}_{3} \mathrm{O}_{7-y} $ crystals measured by Ito et al. \cite{34'}.
	Theoretical curves, inverse of Eq.~\eqref{}, fit the data extremely precisely. 
	Labels show $7-y$ values.
	}
	\label{fig15'}
\end{figure}

In order to examine the difference of effectiveness between $\rho(T)$ and $\sigma(T)$ for analysing the experimental data, we take the data of $ \mathrm{YBa}_{2} \mathrm{Cu}_{3} \mathrm{O}_{7-y} $ crystals by Ito et al. \cite{34'}; 
their results for $y=0.55$, $0.42$, $0.32$ and $ 0.22$ samples are reproduced in Fig.~\ref{fig15'}. 
First, we have analysed $ \rho (T) $ for the temperature range, $100 \; \mathrm{K} \leq T \leq 200 \; \mathrm{K} $ on assuming the form $ \rho (T) = \rho_0 - b T^2 \ln ( T / T^* )$, and found that $\rho_0 $ becomes negative for $ y = 0.22 $ and $ 0.32 $ samples; 
we have concluded that the analysis of $ \rho (T)$ is not very effective. 
Next we have analysed their data, by reading $8$ points for the range $125 \; \mathrm{K} \leq T \leq 300 \; \mathrm{K} $, in terms of the conductivity in the form
\begin{align} \label{58}
\sigma(T) = \sigma_0 \left\{ 1 + \left( \frac{T}{T_1} \right)^2 \ln \frac{ T }{ T^* } \pm \left( \frac{T}{ T_2 } \right)^4 \ln \frac{T}{T^{**}} \right\}
\: , 
\end{align}
where the minus sign of the last term is for $y=0.55$ sample.
It is truly remarkable to find that, if we express the experimental data in $3$-digit numbers, the fitting errors are absolutely zero;
Eq.~\eqref{58} reproduces the data extremely precisely.
The results are listed in Table~\ref{newtab2}.
%
%
%
\begin{table*}[!t]
\caption{Constants appearing in $\sigma(T)$ of $\mathrm{YBa}_{2} \mathrm{Cu}_{3} \mathrm{O}_{7-y}$ crystals, Eq.~\eqref{58}.}
\label{newtab2}
\centering
\renewcommand{\arraystretch}{1.2}
\begin{tabular}{SSSSSS} \hline
 {$y$} &
 {$ \sigma_0 \, ( 10^4 \mathrm{ \Omega^{-1} \cdot cm^{-1} } )$} &
 {$ T_1 \, ( \mathrm{K} )$} &
 {$ T^* \, ( \mathrm{K} ) $} &
  {$ T_2 \, ( \mathrm{K} )$} &
  {$ T^{**} \, ( \mathrm{K} ) $}  \\ \hline
0.22 & 2.36 & 101 & 101 & 191 & 885 \\
0.32 & 2.31 & 98  & 155 & 187  & 871 \\
0.42 & 1.63 & 111 & 173 & 208 & 946 \\
0.55 & 0.750  & 175 & 318 & 646 & 9253 \\ \hline 
\end{tabular}
\renewcommand{\arraystretch}{1}
\end{table*}
It is concluded that the electric conduction of YBCO samples is most properly described in terms of $\sigma(T)$; 
the logarithmic temperature dependence of $ \sigma(T)$ confirms that the principal electric conduction is characterized by the impurity scattering in the 3-dimensional system.

Finally we shall describe the correlation between $T_{\mathrm{C}}$ and $ \rho_1 / \rho_0 $ as variables of composition $x$; i.e. $\rho_1/ \rho_0$ and $ T_{\mathrm{C}} $  vs $x$ diagram.
For that purpose, we have analysed the resistivity data by Nakamura and Uchida \cite{5};
they have measured $ \rho_{ab} (T) $ and $ \rho_{c} (T) $ of high quality single crystals of LSCO for the composition range $ 0.10 \leq x \leq 0.30 $.
They have  analysed the data with the view of the crossover from two to three dimensions, and discussed the scope of $T$-linear and $ T^{n} \, ( n >1)  $ resistivities.
We have reexamined their data and found that their resistivities can be precisely fitted by the $ \rho_0 - \rho_1 T^2 \ln ( T/ T_{\rho}^{*} ) $ law.
%


\begin{figure}[!t]
	\centering
    \includegraphics[width=\linewidth]{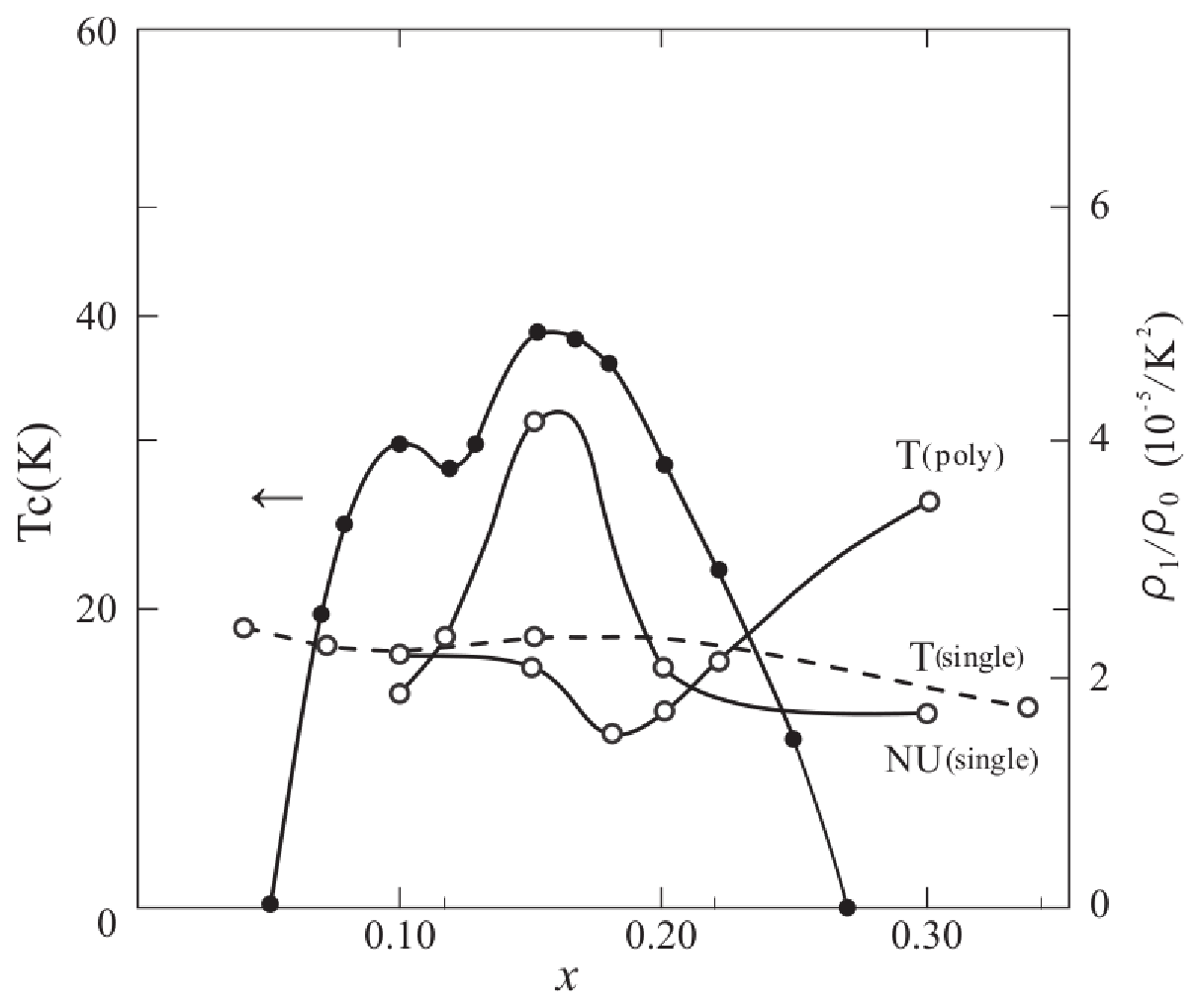}
	\caption{
	$\rho_1 / \rho_0 $ and $T_{\mathrm{C}}$ vs $x$ diagram of $\mathrm{La}_{2-x} \mathrm{Sr}_{x} \mathrm{CuO}_4 $ single crystal and polycrystalline samples.
	T (poly or single) or NU (single) denotes the experimental data of Takagi et al \cite{26} or of Nakamura and Uchida \cite{5}.
	Lines are guides to the eye.
	}
	\label{fig12}
\end{figure}


In Fig.~\ref{fig12}, we plot the $ \rho_1 / \rho_0 $ and $ T_{\mathrm{C}} $ vs $x$ diagram on the basis of Takagi et al's experiment (T) and Nakamura and Uchida's experiment (NU).
The NU data show a strong correlation between $ \rho_1 / \rho_0 $ and $ T_{\mathrm{C}}$,  while the T data both for single crystal and polycrystalline samples do no correlation at all.
To clarify how the nature or configuration of impurities and/or imperfections influence the $ \rho_1 / \rho_0 $ values is a future problem.
Thus we conclude that microscopic theories based on the first principle seem to be ineffective at present.

In the literature on the theories of electrical resistivity for normal-state cuprates, most authors treat them on the basis of the electron-electron scattering in 2D systems; these treatments, however, do not produce the $T^2 \ln T$ dependence.

\section{Hall Coefficient of LSCO and NCCO}


\begin{figure}[!hbt]
	\centering
    \includegraphics[width=\linewidth]{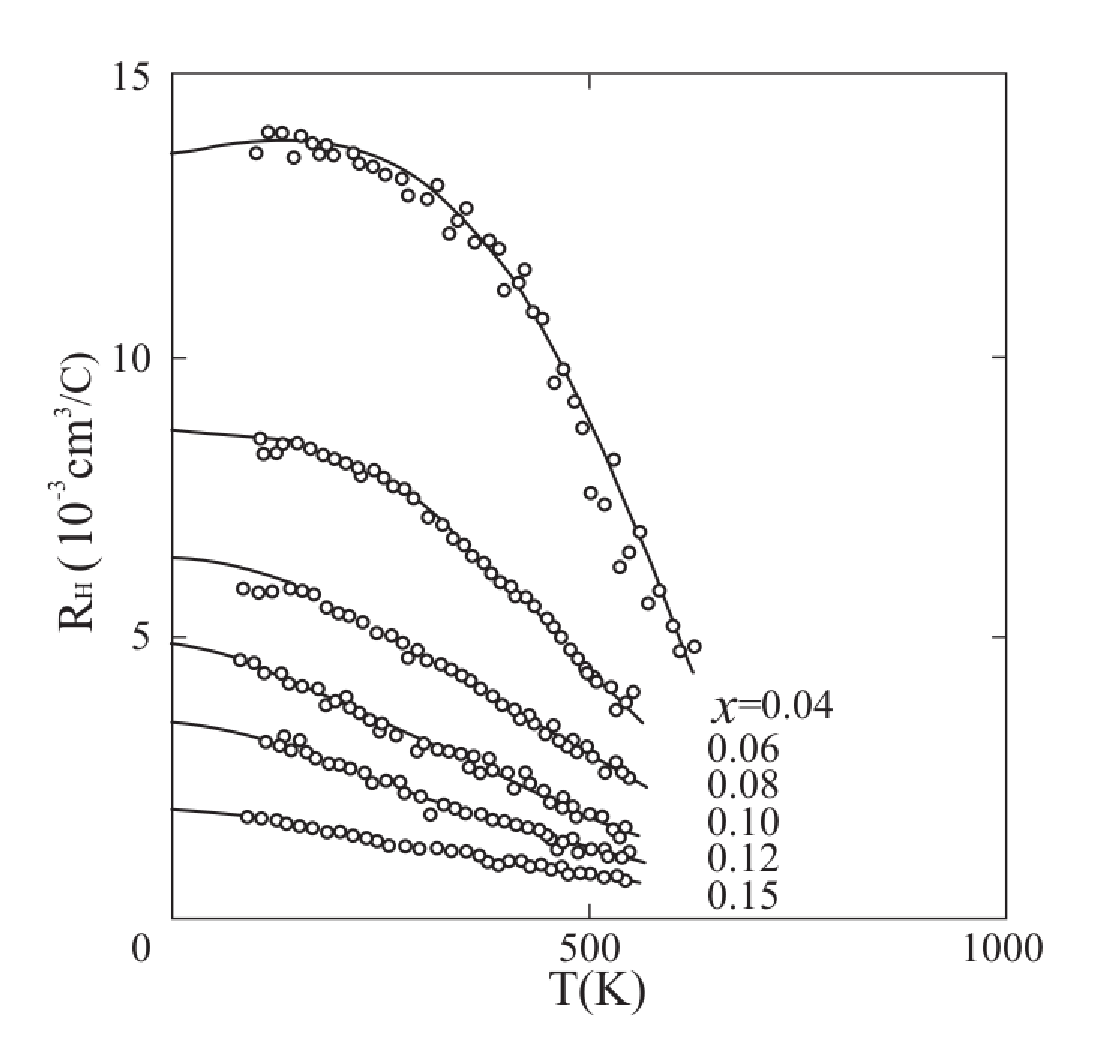}
	\caption{
	$R_{\mathrm{H}} (T) $ vs $T$ of $\mathrm{La}_{2-x} \mathrm{Sr}_{x} \mathrm{CuO}_{4} $ samples.
	Circles show the experimental data \cite{12}; solid lines are theoretical fits, Eq.~\eqref{7.1} in the text.
	}
	\label{fig13}
\end{figure}


Nishikawa et al. \cite{12} have performed extensive experimental studies on transport properties of LSCO sintered samples; the results for the Hall coefficient, $ R_{\mathrm{H}} (T) $, are reproduced in Fig. \ref{fig13}.
Facing these data, they describe the temperature variation of $ R_{\mathrm{H}}$ as a drastic change or crossover like change.
Standing on the view that $R_{\mathrm{H}}$ is inversely proportional to the carrier number density $p$, they state that the temperature change of $ R_{\mathrm{H}} $ of $ x = 0.04 $ sample exhibits the crossover from the low-$p$ to high-$p$ state.

Here we shall describe the $ R_{\mathrm{H}} (T)  $ vs $T$ curves on the basis of the $ T^2 \ln T $ variation.
Six curves of the experimental data, from $ x=0.04$ to $0.15$ sample, can be fitted to the formula
\begin{align} \label{7.1}
R_{\mathrm{H}} (T)  =  R_0 \left\{ 1 \mp \left( \frac{T}{T_1} \right)^2 \ln \frac{ T }{ T_{R}^*}  \right\}
\: , 
\end{align}
where $R_0$, $T_1$ and $T_R^*$ are constant.
As seen from the Figure, all theoretical curves pass through the midpoints within the scattering the experimental data.
For $x=0.04 $ sample, we obtain 
\setcounter{equation}{58}
\begin{subequations}
\begin{align} 
R_{\mathrm{H}} (T)  =  13.9 \left\{ 1 - \left( \frac{T}{781} \right)^2 \ln \frac{ T }{ 207 }  \right\} \times 10^{-3} \mathrm{cm^3/C}  \: .
\end{align}
At $x=0.067$, the sign of the $ T^2 \ln T $ term changes; 
for $x=0.08$ sample, 
\begin{align}
R_{\mathrm{H}} (T)  =  6.44 \left\{ 1 + \left( \frac{T}{938} \right)^2 \ln \frac{ T }{ 3321 }  \right\} \times 10^{-3} \mathrm{cm^3/C}  \: .
\end{align}
Finally, for $x=0.15$ sample, we obtain 
\begin{align}
R_{\mathrm{H}} (T)  =  1.92 \left\{ 1 + \left( \frac{T}{547} \right)^2 \ln \frac{ T }{ 960 }  \right\} \times 10^{-3} \mathrm{cm^3/C}  \: .
\end{align}
\end{subequations}
Since $R_0$, which is expressed by 
\begin{align}
R_0 = 31.25 \left\{ 1 + 15.74 x + 420 x^2  \right\}^{-1} \times 10^{-3} \mathrm{cm^3 / C} \: , 
\end{align}
and the magnitude of the $T^2 \ln T$ term are smoothly varying function of $x$, 
we may not find the crossover effect in $ R_{\mathrm{H}} (T)$.
%


\begin{figure}[!ht]
	\centering
    \includegraphics[width=\linewidth]{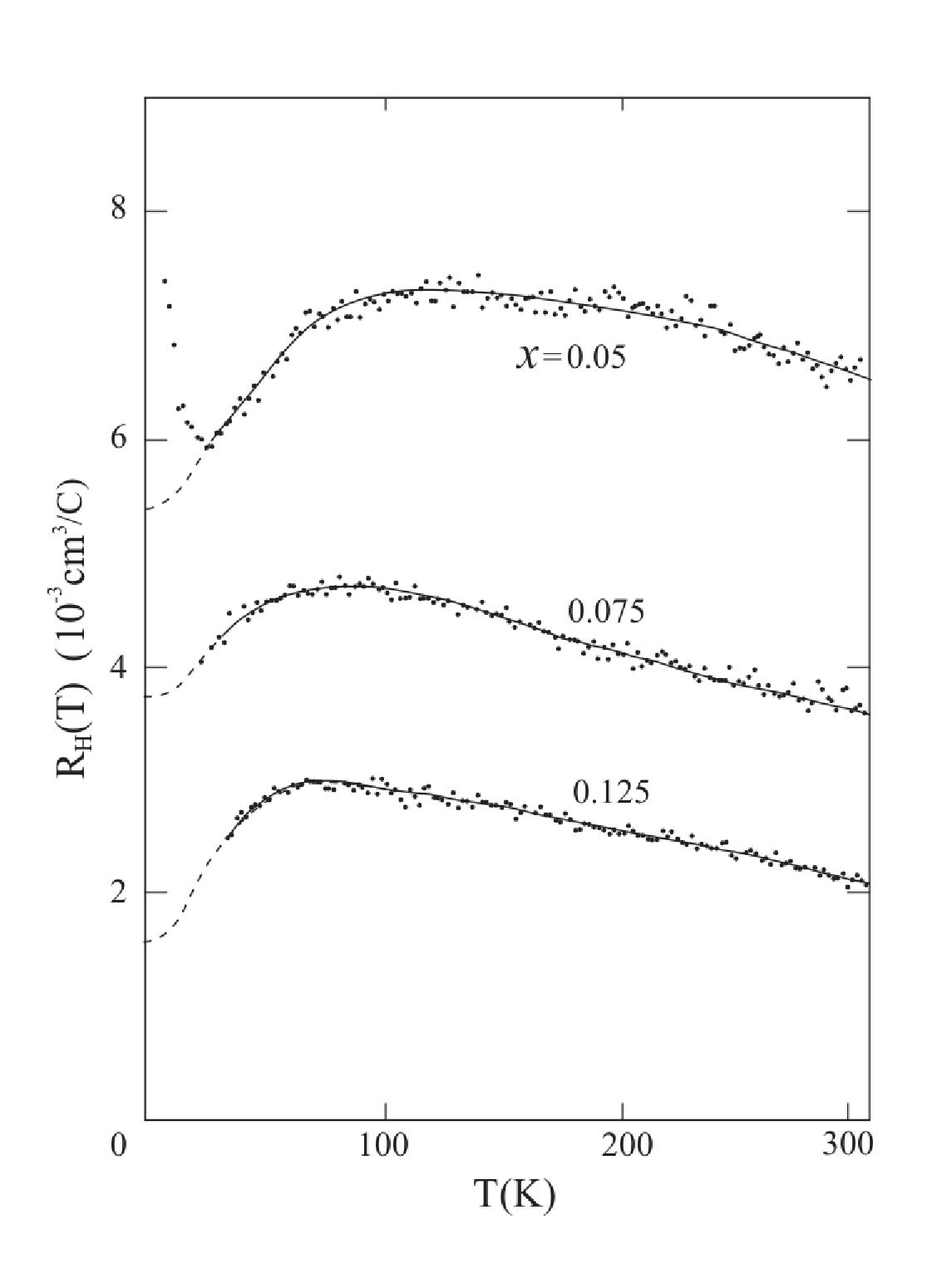}
	\caption{$R_{\mathrm{H}} (T) $ vs $T$ of $\mathrm{La}_{2-x} \mathrm{Sr}_{x} \mathrm{CuO}_{4} $ samples.
	Dots show the experimental data \cite{28};
	solid and dashed lines show the theoretical fit, Eq.~\eqref{7.3} in the text.
	}
	\label{fig14}
\end{figure}


We have to take notice that the experimental data are very sensitively influenced by the sample preparation conditions.
In Fig. \ref{fig14}, the experimental data of $R_{\mathrm{H}}$ for LSCO polycrystalline samples by Hwang et al. \cite{28} are presented.
Here, for low $x$ samples, $R_{\mathrm{H}}$ does not vary significantly for the temperature range from $ 50 \, \mathrm{K}$ to $ 300 \, \mathrm{K} $; a steep drop as seen in $ x= 0.04 $ sample by Takeda et al. does not appear here.
Because of slow change with $T$, here, we should fit the data by considering higher order terms;
\begin{align} 
R_{\mathrm{H}} (T)
\; &= \; R_0 \,\bigg\{
	1
	- \bigg( \frac{ T}{T_1} \bigg)^2 \ln \frac{ T }{ T_1^* }
	- \bigg( \frac{ T }{ T_2 } \bigg)^4   \ln \frac{ T }{ T_2^* }
	\notag 
	\\
& \qquad  \:\: \quad 
	+   \bigg( \frac{ T }{ T_3 } \bigg)^4 ( \ln T )^2   \bigg\} \: ,
	\label{7.3}
\end{align}
where, for the change of $x$ from $0.05$ to $0.125$, constants $R_0, \; T_1, \; T_1^* ,\; T_2, \; T_2^*$ and $T_3$ change smoothly from $5.40$ to $1.55$, from $86.7$ to $37.0$, from $79.8$ to $57.6$, from $86.5$ to $49.7$, from $29.3$ to $26.1$, and from $165$ to $86.5$.
It should be remarked that, plotting these 6 constants as functions of $x$, all curves are found to be almost linear functions.


\begin{figure}[!t]
	\centering
    \includegraphics[width=\linewidth]{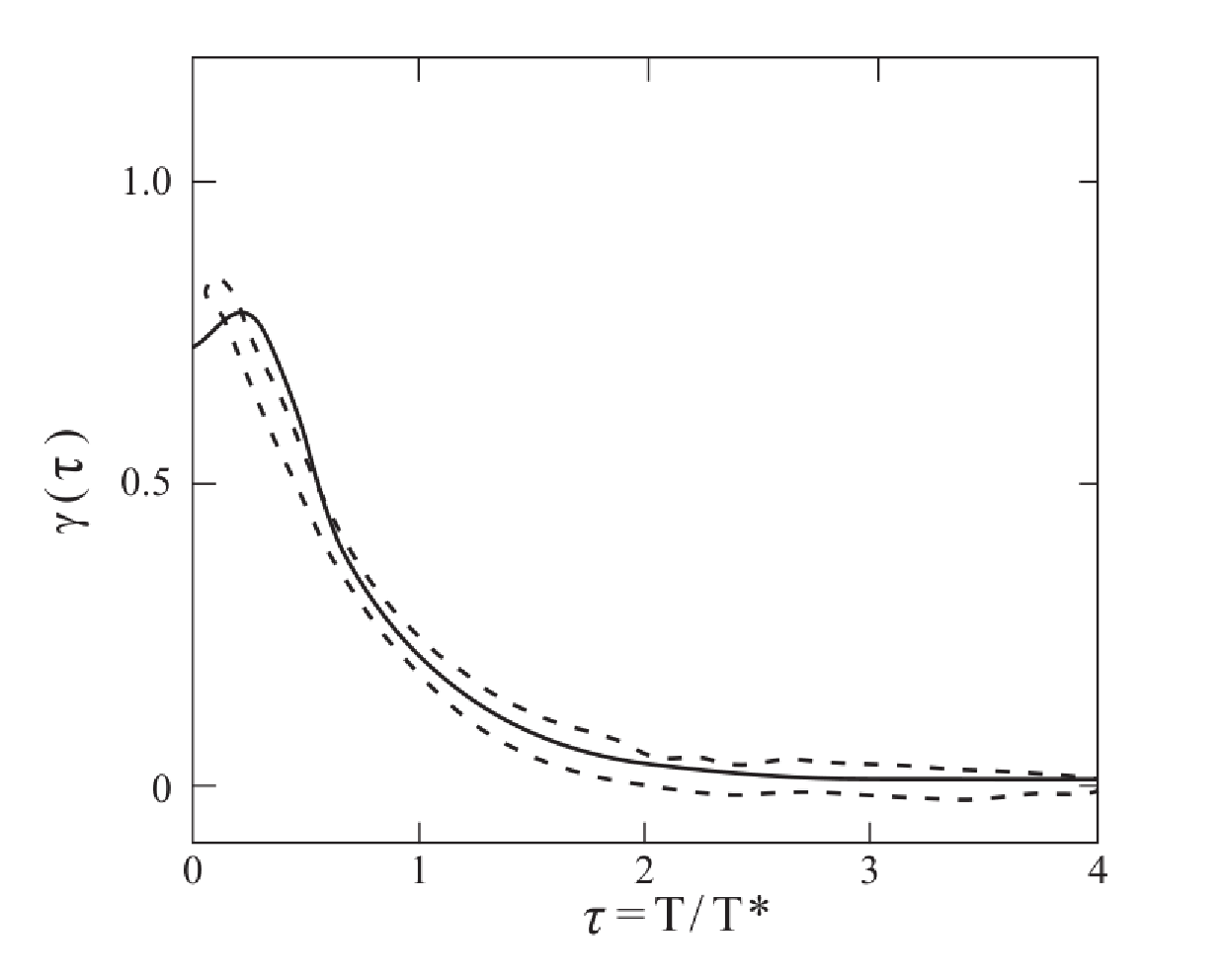}
	\caption{
	The universal curve $\gamma (\tau)$ for the Hall coefficient of LSCO samples as a function of $\tau = T/ T^*$ determined by Hwang et al \cite{28}.
	All experimental points are densely populated in the region surrounded by two dashed lines; the solid line shows the theoretical fit, Eq.~\eqref{7.4} in the text.
	}
	\label{fig15}
\end{figure}

Hwang et al., on the basis of preceding data, have remarked the existence of a universal law for $R_{\mathrm{H}} (T)$; all $R_{\mathrm{H}} (T)$ curves of single and polycrystalline samples and of $0.15$ to $0.34$ compositions can be expressed in a single relation.
Here, as $T$ increases, $R_{\mathrm{H}} (T)$ approaches its asymptotic value $R(\infty)$; $T^*$ is defined approximately as the temperature of the crossover from $T$-dependent to $T$-independent $R (T)$; $R^*$ is a measure to renormalize $R_{\mathrm{H}} (T)$.
Then, all $R_{\mathrm{H}} (T)$ data can be expressed in a single formula $ R(\infty) + R^* \gamma (T/T^* ) $, where $\gamma (\tau)$ represents the universal law; their $\gamma (\tau) $ data are reproduced in Fig.~\ref{fig15}.
Facing to this fact, however, they have stated, ``A simple mathematical expression describing $\gamma (\tau)$ could not be found.
For example, the expression $ A + B / ( \tau + \theta) $ is not adequate''.
Now, we have found that $\gamma(\tau)$ can be expressed by
\begin{align} 
\gamma( \tau) \; &= \; \frac{  R_{\mathrm{H}} (\tau ) - R_{\mathrm{H}} (\infty)  }{ R_{\mathrm{H}}^* }
\notag 
\\
\label{7.4}
 &= \; 0.731 \, \bigg\{  1 + \bigg(  \frac{ \tau}{0.705} \bigg)^2 \ln \frac{ \tau }{ 0.317 }  \bigg\}^{-1}
\: .
\end{align}
This $ \gamma(\tau) $ pass through, in Fig. \ref{fig15},
the midpoints of the scattering of the experimental data.
This form of $\gamma(\tau)$ symborizes  the effectiveness of logarithmic functions for describing transport properties.


\begin{figure}[h]
	\centering
    \includegraphics[width=\linewidth]{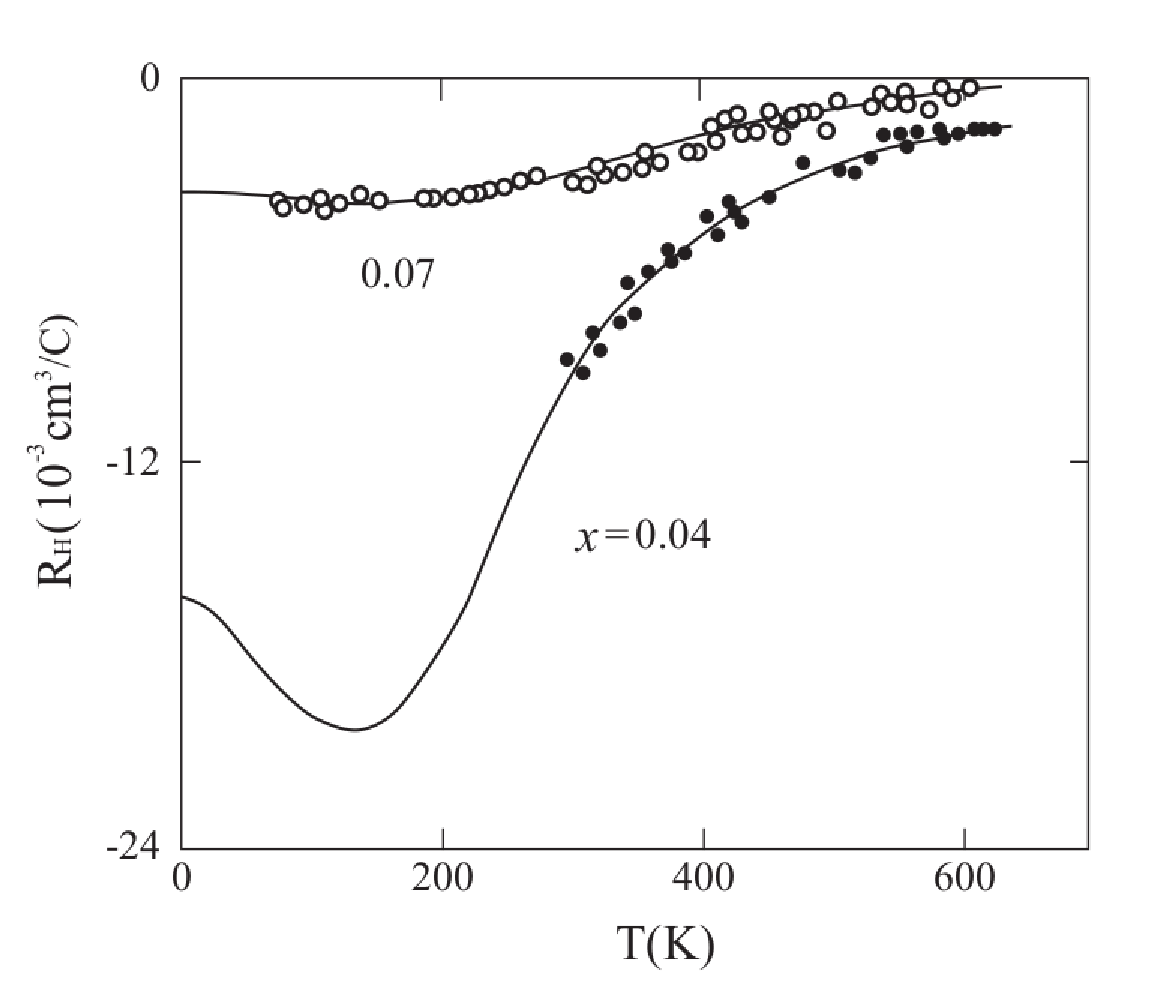}
	\caption{
	$ R_{\mathrm{H}} (T) $ vs $T$ of $\mathrm{Nd}_{2-x} \mathrm{Ce}_{x} \mathrm{CuO}_{4}  $ samples.
	Circles show the experimental points \cite{12}, and solid lines are theoretical fits, Eqs. \eqref{7.6} and \eqref{7.7}, which are extended up to zero temperature for reference.
	}
	\label{fig16}
\end{figure}


As an example of showing negative values of $R_{\mathrm{H}} (T)$, we shall analyse the data of $\mathrm{Nd}_{2-x} \mathrm{Ce}_{x} \mathrm{CuO}_4$ samples measured by Nishikawa et al. \cite{12}; $R_{\mathrm{H}} (T)$ for $x= 0.07$ and $0.04$ samples is reproduced, because of the lesat experimental scattering in Fig. \ref{fig16}.
We have found that, while the $ (a - b T^2 \ln T) $ law is valid for $ R_{\mathrm{H}} (T) $ up to $ 300 \; \mathrm{K} $, the same law is effective for $ 1/ R_{\mathrm{H}} (T) $ up to $ 600 \; \mathrm{K} $. Thus, $ R_{\mathrm{H}} (T) $ can be fitted by
\begin{align} \label{7.6}
1/ R_{\mathrm{H}} (T) = - 0.0615 \bigg\{  1 + \bigg( \frac{ T }{ 208 } \bigg)^2  \ln \frac{ T }{ 215 }  \bigg\} \; \mathrm{cm}^{-3} \cdot \mathrm{C} \: ,
\end{align}
for $ x=0.07 $, and
\begin{align} \label{7.7}
  1/ R_{\mathrm{H}} (T)
  =
  - \, 0.327 \bigg\{
                1 + \bigg( \frac{ T}{251} \bigg)^2 \ln \frac{ T }{ 300 }
                  \bigg\}  \: \mathrm{cm}^{-3} \! \cdot \! \mathrm{C} \: ,
\end{align}
for $x = 0.07$.
Here, for $x= 0.04$ sample, the $ R_{\mathrm{H}}  $ minimum appears at $ 145 \; \mathrm{K} $ with $ R_{\mathrm{H}} ( 145 \, \mathrm{K} ) = - 3.90 \times 10^{-3} \; \mathrm{cm}^3 / \mathrm{C} $.
According to Eq.~\eqref{7.6}, for $x= 0.04$ sample, we predict that the $R_{\mathrm{H}}$ minimum appears at $ 130.6 \; \mathrm{K} $ with $ R_{\mathrm{H}} = -\,20.3 \times 10^{-3} \; \mathrm{cm}^{-3} / \mathrm{C} $;
if the minimum is really observed, this confirms the validity for the law of $ T^2 \ln T $ variation.

\section{Thermoelectric Power of LSCO}



\begin{table}[!t]
\caption{Constants concerning the thermoelectric power of LSCO, Eq.~\eqref{8.1}}
\label{tab2}
\centering
\renewcommand{\arraystretch}{1.2}
\begin{tabular}{cccc} \hline
 $x$ &
 $ q \, (\mathrm{ \mu V / K^2 } )$ &
 $ T_1 \, ( \mathrm{K} )$ &
 $ T_{Q}^* \, ( \mathrm{K} ) $ \\ \hline
 0.10 & 0.603 & 345 & 582  \\
 0.15 & 0.514 & 430 & 850  \\
 0.20 & 0.360 & 360 & 645  \\
 0.30 & 0.261 & 279 & 506  \\ \hline
\end{tabular}
\end{table}
\renewcommand{\arraystretch}{1}

\begin{table*}[!tb]
\caption{Constants concerning the thermoelectic power of LSCO, Eq.~\eqref{8.2}. NU or ZG stands for Nakamura and Uchida or Zhou and Goodenough.}
\label{tab3}
\centering
\renewcommand{\arraystretch}{1.2}
\begin{tabular}{ccccccc} 
\hline
 Author &
 $x$ &
 $ q \, (\mathrm{ \mu V / K^2 } )$ &
 $ T_1 \, ( \mathrm{K} )$ &
 $ T_1^* \, ( \mathrm{K} )$ &
 $ T_2 \, ( \mathrm{K} )$ &
 $ T_2^* \, ( \mathrm{K} )$ \\ \hline
 NU & \begin{tabular}{c} 0.10 \\ 0.15 \\ 0.20 \end{tabular}
 	& \begin{tabular}{c} 0.863 \\ 0.890 \\ 0.497 \end{tabular}
 	& \begin{tabular}{c} 126 \\ 96 \\ 97 \end{tabular}
 	& \begin{tabular}{c} 166 \\ 144 \\ 147 \end{tabular}
 	& \begin{tabular}{c} 211 \\ 177 \\ 179 \end{tabular}
 	& \begin{tabular}{c} 826 \\ 798 \\ 799 \end{tabular}  \\  \hline
 ZG & \begin{tabular}{c} 0.15 \\ 0.20 \\ 0.21\end{tabular}
 	& \begin{tabular}{c} 0.736 \\ 0.352 \\ 0.307  \end{tabular}
 	& \begin{tabular}{c} 100 \\ 113  \\ 109 \end{tabular}
 	& \begin{tabular}{c} 141 \\ 170  \\ 166 \end{tabular}
 	& \begin{tabular}{c} 177 \\ 206 \\ 201 \end{tabular}
 	& \begin{tabular}{c} 749 \\ 894 \\ 889\end{tabular}  \\ \hline
  \end{tabular}
\renewcommand{\arraystretch}{1}
\end{table*}

Zhou and Goodenough \cite{29} have performed intensive analysis for characteristic natures of thermoelectric power $Q (T)$.
As conclusions, they have stated that the temperature and composition dependence of $Q(T)$ of cuprate superconductors cannot be understood within the framework of electron-phonon interactions, the marginal-Fermi liquid theory, or any of the several magnetic models which have been proposed to describe the electronic properties of cuprate-superconductors.
In the subsequent paper \cite{40}, they have reached the conclusion;
there is little evidence that magnetic or electron-electron interactions alone or the location of a Fermi energy in a van Hove singularity is responsible for the peculiar transport properties of the normal state of cuprate superconductors.
To the contrary, we have shown that $Q(T)$ of the 3D Fermi-liquid should universally exhibit the maximum or minimum as a function of temperature \cite{2,13}.


\begin{figure}[!bh]
	\centering
    \includegraphics[width=\linewidth]{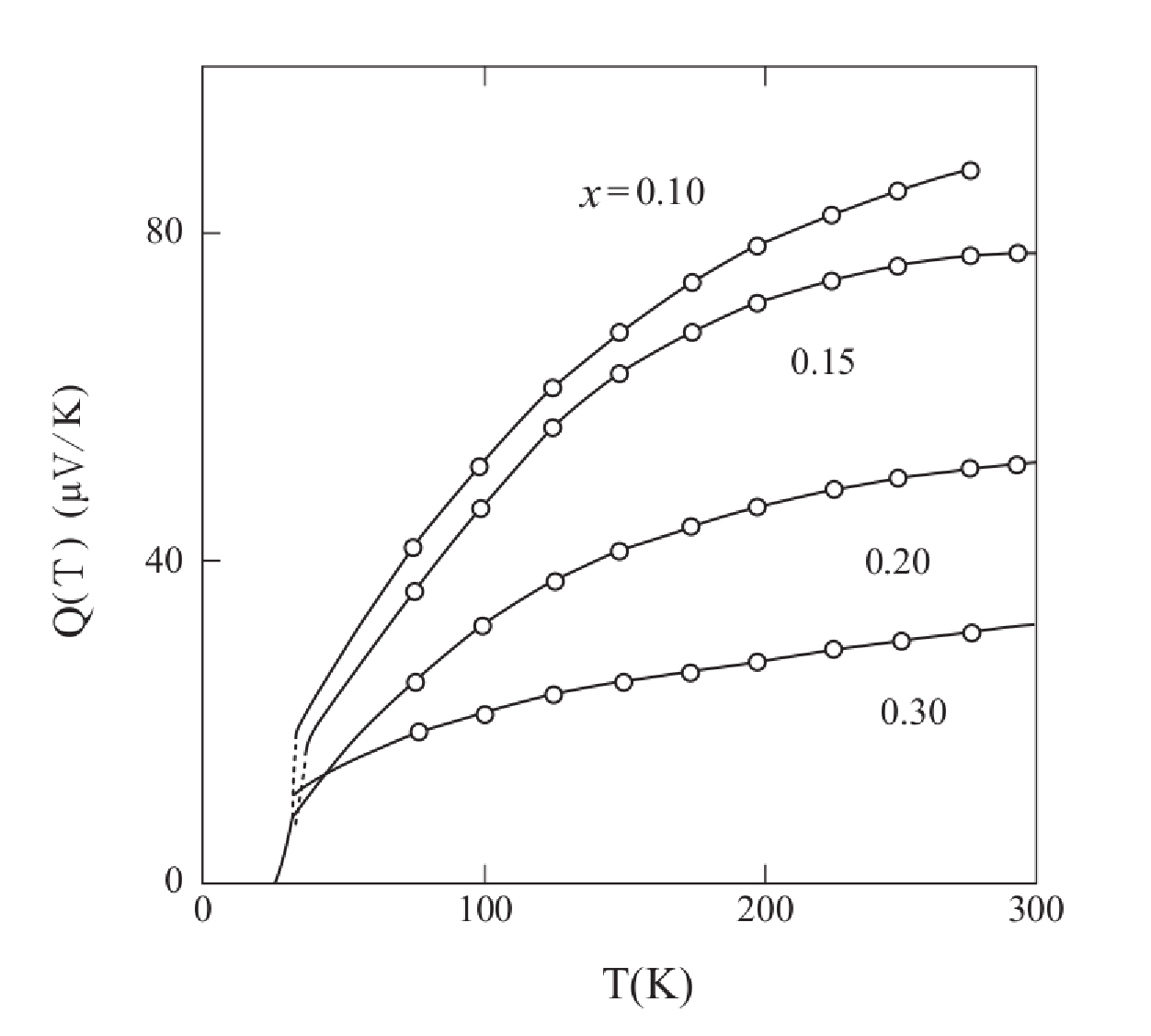}
	\caption{
	$Q_{c} (T) $ vs $T$ of $\mathrm{La}_{2-x} \mathrm{Sr}_x \mathrm{CuO}_4$ single crystals.
	Solid lines show the experimental data \cite{5}; circles show the representative points of the theoretical fit, Eq.~\eqref{8.1} in the text.
	}
	\label{fig17}
\end{figure}


Nakamura and Uchida \cite{5} have measured the thermoelectric power, $Q_{ab} (T)$ and $Q_c (T)$, of LSCO high quality single crystals.
As shown in Fig. \ref{fig17}, $Q_c(T)$'s are monotonically increasing functions of $T$ up to about $300 \, \mathrm{K}$.
The data can be precisely fitted by the formula
\begin{align} \label{8.1}
Q_c (T) = q T \bigg\{  1 + \bigg( \frac{ T}{T_1}\bigg)^2 \ln \frac{T}{T_Q^*}  \bigg\} \: .
\end{align}
%



\begin{figure}[!h]
	\centering
    \includegraphics[width=\linewidth]{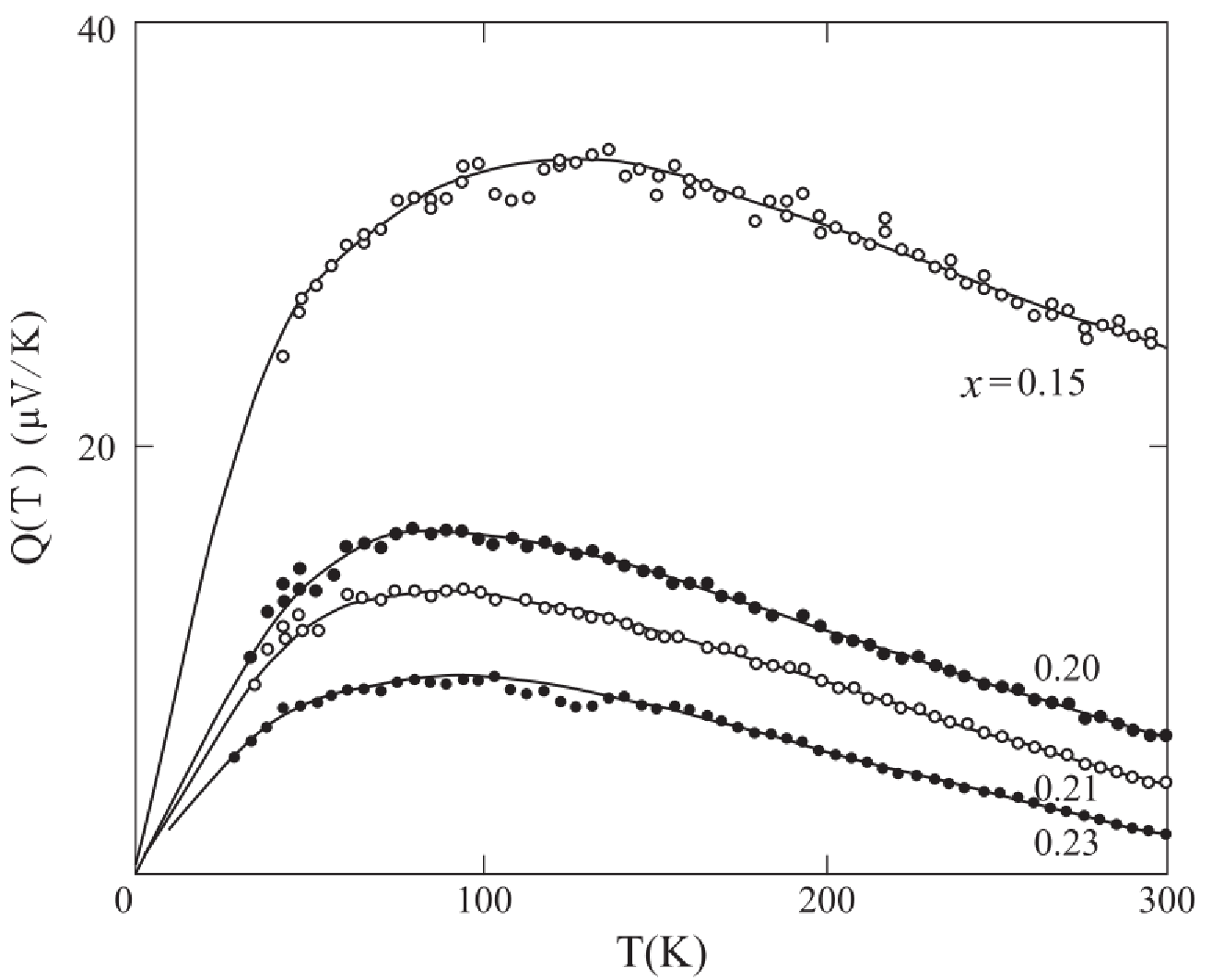}
	\caption{
	$Q_{ab} (T) $ vs $T$ of $\mathrm{La}_{2-x} \mathrm{Sr}_{x} \mathrm{CuO}_{4} $ samples.
	Circles show the experimental data \cite{29}; solid lines show the theoretical fit, Eq.~\eqref{8.2} in the text.
	}
	\label{fig18}
\end{figure}

\noindent
The constants $q$, $T_1$ and $T_Q^*$ are listed in Table \ref{tab2};
these are found to be smoothly varying functions of $x$.
The average $q$-value for LSCO is about $0.5 \, \mathrm{\mu V / K^2}$;
this value is half of that of iron-based superconductors, and $1/60$  of that of $\mathrm{Ce Al_3}$.


$Q_{ab} (T)$, on the other hand, shows the maximum in the vicinity of $100 \, \mathrm{K}$ and then decreases with $T$.
For reference, we compare their data with the data by Zhou and Goodenough \cite{29} regarding the same $Q_{ab} (T)$; the latter is reproduced in Fig. \ref{fig18}.
We have found that, since $Q_{ab} (T) $'s data are slowly varying functions of $T$, we need to consider next higher order term of $T^5 \ln T$;
\begin{align} \label{8.2}
Q_{ab} (T) = qT \bigg\{ 1 + \bigg( \frac{T}{ T_1 } \bigg)^2 \ln \frac{ T }{T_1^*} + \bigg( \frac{T}{T_2} \bigg)^4 \ln \frac{ T }{ T_2^* } \bigg\} \: .
\end{align}
Both NU's and ZG's data can be nicely fitted by Eq.~\eqref{8.2}.
The fitted results are listed in Table \ref{tab3}.

On observing the table, we find two remarkable facts; first, although the absolute values of $Q_{ab}$ for NU $0.15(x)$ and $0.20 (x)$ samples are quite different, the relative temperature variations are almost exactly the same; secondly, the relative temperature variations of NU $0.15$ and ZG $0.15$ samples are almost the same.
This problem is related with sample preparation conditions, characters of impurities or imperfections, the existence of a universal law, etc., and await future intensive researches.

\section{Conclusions}

We have shown that the quasiparticles in the normal state of high-$T_{\mathrm{C}}$ cuprate superconductors behave as a 3D Fermi liquid. %
Because of interactions and the presence of Fermi energies, the thermodynamic and transport properties obey the logarithmic formula with respect to temperature $T$.
The electronic specific heat coefficient, magnetic susceptibility, electrical resistivity, Hall coefficient and thermoelectric power divided by temperature follow the formula, $a- b T^2 \ln (T/T^*)$, $a$, $b$ and $T^*$ being constant.
This formula explains observations for the normal state of LSCO, YBCO and NCCO compounds.
The kink phenomenon which appears in $\mathrm{Re} \, \Sigma( \varepsilon)$ is shown to be attributed  to its $\varepsilon^3 \ln | \varepsilon |$ dependence.
The appearance of the spin gap is not required for explaining the $T$-dependence of the magnetic susceptibility.
On deriving above formulae, quasiparticles in cuprates are assumed to be 3-dimensional; this has been surely confirmed in the specific heat data.
The $T^2 \ln T$ variation of transport properties arises exclusively from the scattering of quasiparticles with impurities in 3D, but does not from the electron-electron scattering in 2D.
Since LSCO has been shown to be neither nearly ferromagnetic nor nearly antiferromagnetic, for the origin of superconductivity, the effect of charge fluctuations and that of spin fluctuations are equally important.

\section*{Acknowledgement}

The author would like to thank S. Okano for careful reading of manuscripts and aid for making the paper refined.

\section*{Statements and Declarations}

The author has no competing interests to declare that are relevant to the content of this article.

\end{document}